\documentclass[pra,twocolumn,superscriptaddress,nofootinbib,nolongbibliography]{revtex4-2}
\usepackage{dcolumn}
\usepackage{bm}
\usepackage{graphicx}
\usepackage{amsfonts}
\usepackage{bbm}
\usepackage{dsfont}
\usepackage{float}
\usepackage{graphicx}
\usepackage{bm}
\usepackage{amssymb}
\usepackage{multibib}
\usepackage{color}
\usepackage{enumerate}
\usepackage{amsmath}
\usepackage{amstext}
\usepackage{latexsym}
\usepackage{braket}
\usepackage{appendix}
\usepackage{textcomp}
\usepackage[usenames,dvipsnames]{xcolor}
\usepackage[colorlinks=true,citecolor=Blue,linkcolor=RubineRed,urlcolor=Blue]{hyperref}

\newcommand{\ZY}{\color{Black}}
\newcommand{\R}{\color{Black}}

\begin{document}

\title{Quantum speed limits in arbitrary phase spaces}
\date{\today}

\author{Weiquan Meng}
\author{Zhenyu Xu}
\affiliation{School of Physical Science and Technology, Soochow University, Suzhou 215006, China}

\begin{abstract}

Quantum speed limits (QSLs) provide an upper bound for the speed of evolution of quantum states in any physical process. {\ZY Based on the Stratonovich-Weyl correspondence,} we derive a universal QSL bound in arbitrary phase spaces that is applicable for both continuous variable systems and finite-dimensional discrete quantum systems. {\ZY This QSL bound allows the determination of speed limit bounds in specific phase spaces that are tighter than those in Wigner phase space or Hilbert space under the same metric, as illustrated by several typical examples, e.g., a single-mode free field and $N$-level quantum systems in phase spaces.} This QSL bound also provides an experimentally realizable way to examine the speed limit in phase spaces relevant to applications in quantum information and quantum optics.

\end{abstract}

\maketitle

\section{Introduction}
Quantum speed limits (QSLs) set the upper bound on the speed for quantum systems evolving in an arbitrary physical process  \cite{Pfeifer2005RMP,BookBusch,Deffner2017Review}. The milestone of describing the QSL time as the intrinsic time scale of quantum dynamics was achieved by Mandelstam and Tamm, who clarified the longstanding debate on the explanations for the time-energy uncertainty relationship \cite{MTbound1945}. The energy fluctuation restricts the minimum time for a quantum state to evolve into its orthogonal state. Apart from the standard deviation of the energy \cite{MTbound1945,Bhattacharyya_1983,AA1990PRL}, alternative QSL time bounds based on the averaged energy \cite{TLbound1998,Luo2005,Magdalena2006PRA,Norman2011} and the interplay among them \cite{Luo2004,Toffoli2009PRL,Ness2022PRL} have been developed in succession. In the last decade, QSLs have even been extended to the fields of open systems involving various metrics \cite{Taddei2013PRL,Adolfo2013PRL,Deffner2013PRL,Zhang2014,Lidar2015PRL,Pires2016PRX,Xu_2016}, quantum dynamical speedup \cite{Frowis2012,Deffner2013PRL,Zhenyu2014PRA,Zhang2015PRA,Liu2016PRA,Cai2017PRA,An2023PRA}, information geometry \cite{Campaioli2018PRL}, time-optimal quantum control \cite{Carlini2006PRL,Caneva2009PRL,Hegerfeldt2013PRL,Wang2015PRL,Du2016PRL}, quantum and macroscopic stochastic processes \cite{Shiraishi2018PRL,LuisP2019NJP,Hamazaki2022PRXQ}, non-Hermitian systems \cite{Adolfo2013PRL,Sun2019PRL,Zheng2021PRL,Impens2021PRA}, dynamics of many-body quantum systems \cite{Campbell2020PRL,Takahashi2020PRR}, the evolution of observables \cite{LuisP2022PRX,Mohan2022PRA,Carabba2022}, and recently {\R the topological structure of the dynamics \cite{Tan2023PRL}}. Indeed, the QSL represents a powerful tool for evaluating the performance of quantum computers \cite{lloyd2000}, the accuracy of quantum metrology \cite{Giovannetti2011,Barbieri2022PRXQuantum}, and the efficiency in quantum thermodynamics \cite{Campaioli2017PRL,Campbell2017PRL,Funo2017PRL,Xu2018PRA,Nicholson2020,Falasco2020PRL,LuisP2020PRL,VanVu2021PRL,Dahlsten2021PRL}, among others. Thus, extensive effort has been applied toward experimental verification to this end \cite{Cimmarusti2015,Alberti2021SA,Adolfo2021PRL,Alberti2021PRX,Feng2022PRL}. 

Apart from the well-known Schr$\ddot{\mathrm{o}}$dinger's wave function, Heisenberg's matrix mechanics, and Feynman's path integral, the phase-space formalism of quantum mechanics has an advantage in terms of its resemblance to classical statistical mechanics and supports a profound understanding of the underlying statistical properties \cite{bookPS}. {\ZY The pioneering work regarding the study of QSLs in phase spaces was presented by Deffner, who found that the QSL in the Wigner ($W$) phase space is equivalent to that in the density operator space and much easier to perform computations involving continuous variable systems \cite{Deffner2017NJP}. Later,} because the $W$ phase space retains the structure of the classical phase space through quantum-classical correspondence, {\ZY Shanahan \textit{et al.} discovered that} the QSL is not a particular quantum phenomenon but a universal property of the time evolution of continuous variable systems in the $W$ phase space \cite{Shanahan2018}. {\ZY Meantime, questions regarding the speed limit in the $W$ phase space have led to the intensive study of the classical speed limit over the past few years  \cite{Shanahan2018,Okuyama2018,BolonekLason2021,Poggi2021PRXQ}. 
}

{\ZY For continuous variable systems, the $W$ function is the most widely used phase-space quasiprobability representation with symmetric ordering of the position and momentum operators (or equivalently $a$ and $a^\dagger$) \cite{Wigner1932}. Nevertheless, it is not unique, and a large number of well-known phase-space quasiprobability distribution functions exist \cite{bookPS}. For instance, the Glauber-Sudarshan $P$ function \cite{Glauber1963,Sudarshan1963} with a normal ordering of $a$ and $a^\dagger$ can be used to diagonalize the density operator in terms of coherent states and is fundamental in photoelectric detection \cite{leonhardt_2010}. Additionally, the Husimi $Q$ function \cite{Husimi1940} with the antinormal ordering of $a$ and $a^\dagger$ provides a possible way to define a nonnegative quasiprobability distribution, and the Cahill-Glauber $s$-parametrized quasiprobability distributions incorporate the above $W$, $P$, and $Q$ functions, with $s=0,1$, and $-1$, respectively \cite{Glauber1969,Cahill1969}.}

In addition, the flourishing development in quantum information and quantum technology {\ZY now requires the consideration of finite-dimensional systems, e.g., an ensemble of spins, to define corresponding quasiprobability distribution functions (see a recent review \cite{Rundle2021}). Commonly, there are two methods. The first method involves generating discrete analogues of the quasiprobability distribution functions \cite{Wootters1987,Heiss2000PRA,Zhu2016PRL}, an approach that has been used in various applications in quantum-state tomography \cite{Leonhardt2004PRL}, resource theory for quantum computation \cite{Ernesto,Veitch_2012,Howard2014}, and studies of the dynamics of many-body quantum systems \cite{Schachenmayer2015PRX,Mink2022PRR}. The second method involves employing generalized coherent states \cite{Agarwal1981PRA,Brif_1998,Brif1999,Luis2004PRA,Tilma_2012,Tilma2016PRL} to provide insights into the study of quantum foundations, e.g., the recent construction of a general statistical framework for finite-dimensional discrete quantum systems \cite{Rundle2019PRA} and the interpretation of quantum entanglement with classical trajectories \cite{Runeson2021PRL}. 
} 

{\ZY At this point, the following two questions naturally arise: (i) Is there a simple QSL bound that can incorporate continuous variable and finite-dimensional quantum systems in arbitrary phase spaces? (ii) What are the tangible benefits for the QSL in other quasiprobability distribution phase spaces compared with those in the $W$ phase space?}

{\ZY To address the above questions,} in this paper, we derive a universal QSL bound for quantum systems in arbitrary phase spaces using the Stratonovich-Weyl (SW) correspondence \cite{Stratonovich1957}. The key concept for the SW correspondence is to transform any operator in Hilbert space into target phase spaces {\ZY ($W$, $P$, $Q$ or others)} with an SW kernel, which is constructible for both continuous variable and finite-dimensional quantum systems in $s$-parametrized phase spaces \cite{Rundle2021}. {\ZY This universal QSL bound has several tangible benefits. (i) First, our QSL bound provides a unified framework for studying the speed limit for continuous variable and finite-dimensional quantum systems in quasiprobability distribution phase spaces. This unified QSL bound is not just for aesthetic appeal but provides a convenient tool to evaluate the speed limit directly in various scenarios.  (ii) Second, our work provides new insight into the search for tighter QSL bounds that are superior to those obtained in the $W$ phase space under the same metric. We use the traditional Cauchy-Schwartz inequality method for the first scaling and the $s$-parametrized phase spaces as tools for secondary scaling, and this latter step is our major contribution. (iii) Third, in the past, phase spaces with $s\neq 0, \pm 1$ were rarely used, as they were usually believed to be less valuable than traditional spaces in quantum optics \cite{leonhardt_2010}. In our paper, we show that these representations also have advantages. The tighter QSL bound achieved with $s\neq 0, \pm 1$ provides evidence that a specific choice of $s$ could also be superior to the commonly used $W$, $P$, and $Q$ phase spaces in studying QSL bounds.}

%{\ZY Based on $s$-parametrized phase spaces, the derived} QSL bound has two components with opposite $s$, named the dual parts.
%
% {\ZY One benefit of such a bound (with a specific choice of $s$) is that it can be tighter than that in the $W$ phase space, which is illustrated in typical examples of a single-mode free field and an $N$-level quantum system in $s$-parametrized phase spaces. In the past, phase spaces with $s\neq 0, \pm 1$ were rarely used, as they were usually believed to be less valuable than traditional spaces \cite{leonhardt_2010}. The tighter QSL bound with $s\neq 0, \pm 1$ provides evidence that the specific choice of $s$ could be superior to that in the commonly used $W$, $P$, and $Q$ phase spaces for studying the QSL bound.}

Our paper is organized as follows. In Sec. II, we derive the universal QSL bound in arbitrary phase spaces. Then, we employ this unified QSL bound to several typical quantum systems, including {\ZY a single-mode free field and an $N$-level quantum system in $s$-parametrized phase spaces} in Sec. III. The experimental consideration is analyzed in Sec. IV. Finally, we summarize in Sec. V with concluding remarks and an outlook of potential applications.

\section{Quantum speed limits in arbitrary phase spaces}
Consider an operator $A$ in Hilbert space; its SW symbol in an arbitrary phase space is \cite{Rundle2021}
\begin{equation}
	F^s_{A}(\eta):=\mathrm{Tr}[A\Delta^s (\eta)],  \label{FA}
\end{equation}
where $\Delta^s (\eta)$ denotes the SW kernel, {\ZY and five criteria (linearity, reality, standardization, covariance, and tracing properties) ensure such a legitimate SW correspondence (see Appendix \ref{SWc}).} $\eta$ is a point in a phase space that determines a state $\ket{\eta}$ ($\eta\rightarrow\ket{\eta}$) in Hilbert space and the index $s$ labels a family of phase spaces, e.g., $F^0_{A}(\eta)$ is the well-known Wigner function \cite{Rundle2021}. Since $A \rightarrow F^s_A (\eta)$ is a one-to-one linear map, it is natural to define the inverse of the SW symbol as 
\begin{equation}
   A=\int d\mu (\eta )F^s_{A}(\eta)\Delta^{-s} (\eta), \label{A}
\end{equation}
where $d\mu(\eta)$ is the invariant integration measure. Then, the trace of the product of the two operators is immediately obtained {\ZY(see Appendix \ref{SWc})}

\begin{equation}
   \mathrm{Tr}(AB)=\int d\mu(\eta) F^s_A (\eta)F^{-s}_B(\eta).  \label{TrAB}
\end{equation}

Consider a quantum system evolving under the Hamiltonian $H$, the overlap between the initial state $\rho_0$ and the final state $\rho_t$ is captured by the relative purity $P_t(\rho_0,\rho_t):=\mathrm{Tr}(\rho_0\rho_t)$ \cite{Toffoli2009PRL,Adolfo2013PRL,Shanahan2018}, which is then transformed by the above SW correspondence
\begin{equation}
  P_{t}(\rho _{0},\rho _{t})=\int d\mu(\eta )F^{-s}_{\rho_{0}}(\eta)F^s_{\rho_{t}}(\eta).\label{P_t}
\end{equation}
As the relative purity is a metric of the state evolution, its time derivative can be regarded as the speed of the state evolution. The primary procedure for obtaining the changing rate of the relative purity is to calculate $\partial_t F^s_{\rho _{t}}(\eta)$, which is achieved by transforming the von Neumann equation $\partial_t \rho _{t}=\frac{1}{i\hbar }[H,\rho _{t}]$ to the phase space
\begin{equation}
        {\ZY \frac{\partial F^s_{\rho _{t}}(\eta)}{\partial t} =\left\{ \left\{F^s_{H},F^s_{\rho _{t}}\right\} \right\}(\eta), } \label{dynamics}
\end{equation}
%\begin{align}
%		\frac{\partial F^s_{\rho _{t}}(\eta)}{\partial t} =&\frac{1}{i\hbar }
%		\mathrm{Tr}[[H,\rho _{t}]\Delta^s (\eta)]  \notag \\
%		=&\left\{ \left\{F^s_{H},F^s_{\rho _{t}}\right\} \right\}(\eta),  \label{dynamics}
%\end{align}
where we employ a generalized Moyal bracket
\begin{equation}
      {\ZY  \left\{ \left\{ F^s_{A},F^s_{B}\right\} \right\} := \frac{1}{i\hbar }( F^s_{A}\star F^s_{B}-F^s_{B}\star F^s_{A}),} \label{Moyal bracket}
\end{equation}
and a generalized star product
\begin{align}
	{\ZY (F^s_{A}\star F^s_{B})(\eta)}
		:=&\int d\mu (\eta ^{\prime })\int
		d\mu (\eta ^{\prime \prime })F^{s^{\prime}}_{A}(\eta ^{\prime })\notag\\
		&\times F^{s^{\prime \prime}}_{B}(\eta ^{\prime \prime })
		\mathrm{Tr}[\Delta^{-s^{\prime}}(\eta ^{\prime })\notag \\
		&\times\Delta^{-s^{\prime \prime }} (\eta ^{\prime \prime })\Delta^s (\eta)].\label{star product} 
\end{align}
Equation (\ref{dynamics}) is also called the generalized Liouville equation \cite{Rundle2021} and governs the dynamics of a phase space function.

{\ZY To this end}, the changing rate of the relative purity is given by
\begin{eqnarray}
	\dot{P}_{t}(\rho _{0},\rho _{t}) &=&\int d\mu (\eta )F^{-s}_{\rho _{0}}(\eta)\left\{ \left\{ F^s_{H},F^s_{\rho _{t}}\right\} \right\}(\eta)  \notag \\
	&=&\int d\mu (\eta )F^{-s}_{\rho _{t}}(\eta)\left\{ \left\{F^s_{\rho _{0}},F^s_{H}\right\} \right\}(\eta). \label{puritychange} \notag\\
\end{eqnarray}
Note that the second line is more convenient for use in experimental verification (see Sec. \ref{EP}) and that the derivation can be found in Appendix \ref{appB}. By means of the Cauchy-Schwarz inequality in \textrm{L}$^2$, i.e., $\left\vert \int f(x)g(x)^{\ast }dx\right\vert ^{2}\leq \int \left\vert f(x)\right\vert ^{2}dx\int \left\vert g(x)\right\vert ^{2}dx $, the norm of the changing rate for the purity is bounded by 
\begin{equation}
    \left\vert \dot{P}_{t}(\rho _{0},\rho _{t})\right\vert \leq V^{s}_{\mathrm{QSL}}(t),
    \label{boundPt}
\end{equation}
with 
\begin{equation}
    {\ZY V^{s}_{\mathrm{QSL}}(t):=\min\left\lbrace \chi_t^{-s}v^s_{\mathrm{QSL}}(0) \;, \; \chi_0^{-s} v^s_{\mathrm{QSL}}(t)\right\rbrace, \label{VQSL}}
\end{equation}
{\ZY where the two terms in $\min \{\cdot,\cdot\}$ are equal if $V^{s}_{\mathrm{QSL}}$ is time-independent.} The two components $\chi^{-s}$ and $v^{s}_\mathrm{QSL}$ of $V^{s}_{\mathrm{QSL}}(t)$ are in dual phase spaces {\ZY with opposite $s$} \cite{Note}.
Here
\begin{equation}
    \chi^s_t:=\left[ \int d\mu (\eta )F^s_{\rho _{t}}(\eta)^{2}\right] ^{\frac{1}{2}},\label{alpha_t}
\end{equation}
and
\begin{equation}
    v^s_\mathrm{QSL}(t):=\left[ \int d\mu (\eta )\left\vert \left\{ \left\{F^s_{\rho _{t}},F^s_{H}\right\} \right\}(\eta)\right\vert ^{2}\right] ^{\frac{1}{2}},\label{vQSL_t}
\end{equation}
where the reality postulate of the SW correspondence is employed in the definition of Eq. (\ref{alpha_t}). Specifically, $\chi^{-s}$ and $v^{s}_\mathrm{QSL}$ in Eq. (\ref{VQSL}) are in a self-dual phase space when $s=0$. Upon integration of Eq.(\ref{boundPt}) with respect to time from $0$ to $\tau$, we obtain the usual definition of QSL time 
\begin{equation}
    \tau \geq \tau _{\mathrm{\mathrm{QSL}}}:=\frac{1-P_{\tau }}{\left\langle V^{s}_{\mathrm{QSL}}(t)\right\rangle _{\tau}}, \label{QSLT}
\end{equation}
where $\left\langle f(t)\right\rangle _{\tau }:=\frac{1}{\tau }\int_{0}^{\tau }f(t)dt$.

This upper bound of the evolution speed $V^{s}_{\mathrm{QSL}}(t)$ is applicable to quantum systems in arbitrary phase spaces as long as the SW kernel is given. The universality of the derived QSL bound lies in the fact that the generalized Moyal bracket and star product are convenient for use with different kinds of systems in various phase spaces, such as continuous and discrete phase spaces, in the same framework. In the following, we examine the derived QSL bound of both continuous variable and finite-dimensional quantum systems in several typical phase spaces, e.g., {\ZY a single-mode of the quantized radiation field in Cahill-Glauber $s$-parametrized quasiprobability distribution phase spaces, $N$-level quantum systems in $s$-parametrized phase spaces with continuous degrees of freedom, and qubit systems in a toroidal lattice phase space with discrete degrees of freedom.}

\section{Examples}
\subsection{A single-mode free field or a one-dimensional harmonic oscillator in phase spaces}\label{subSec3}
For a single-mode free field or a one-dimensional harmonic oscillator, the $s$-parametrized SW kernel is given by \cite{Glauber1969}
\begin{equation}
    \Delta^s(\eta)=\int d\mu (\zeta) D(\zeta) \mathrm{e}^{\eta \zeta^{*}-\eta^{*} \zeta+s|\zeta|^{2} / 2}, \label{DeltaC} 
\end{equation}
defined in terms of the displacement operator $D(\zeta):=e^{\zeta a^\dagger -\zeta^* a}$, and the integration measure $d\mu (\zeta)=(1/\pi) d^2 \zeta$. This $s$-parametrized SW kernel has been widely studied in the field of quantum optics \cite{knight2004,BookKlimov}. Considering the system initially prepared in a coherent state $\ket{\alpha_0}$, the evolved state under the control of $H=\hbar \omega\left(a^{\dagger} a+\frac{1}{2}\right)$ is $\left|\alpha_t\right\rangle=e^{-i \omega t / 2}\left|\alpha_0 e^{-i \omega t}\right\rangle=e^{-i \omega t / 2}\ket{\alpha}$ \cite{knight2004}, where we use $\alpha:=\alpha_0 e^{-i\omega t}$ as an abbreviation. Then, the SW symbol of the coherent state is given by $F^s_{\rho_t}(\eta)=\frac{2}{1-s} \exp \left( \frac{2}{s-1}|\alpha-\eta|^2 \right)$ with $-1\leqslant s<1$ \cite{Cahill1969}. {\ZY In terms of Eqs.(\ref{alpha_t}) and (\ref{vQSL_t})}, we are ready to obtain the two components of $V^{s}_{\mathrm{QSL}}$ (see Appendix \ref{app4}): 
\begin{equation}
    \chi^s=\frac{1}{\sqrt{1-s}}
\end{equation}
and
\begin{equation}
    v^s_{\mathrm{QSL}}=\frac{\sqrt{2} \omega \left|\alpha_0\right|}{1-s},
\end{equation}
which are time-independent ({\ZY thus, we omit the subscript $t$ for notation simplicity}) and lead to the final result 
\begin{equation}
    V^{s}_{\mathrm{QSL}}=\frac{\sqrt{2}\omega |\alpha_0|}{(1-s)\sqrt{1+s}}. \label{VQSL3}
\end{equation}

It is convenient to obtain the QSL bound in Hilbert space $V_{\mathrm{QSL}}=\sqrt{\mathrm{Tr}(|\dot{\rho}_t|^2)}=\frac{\sqrt{2}}{\hbar}\Delta E=\sqrt{2}\omega |\alpha_0|$, which is equivalent to the case of $s=0$ {\ZY in Eq. (\ref{VQSL3}). Thus, Eq. (\ref{VQSL3}) can be rewritten as $V^s_{\mathrm{QSL}}=\frac{\sqrt{2}}{\hbar (1-s)\sqrt{1+s}} \Delta E$ in terms of the standard deviation of $H$. This is a QSL bound of the Mandelstam-Tamm type for the $s$-parametrized phase space, which is in agreement with the QSL bound derived in Ref. \cite{Shanahan2018} in Wigner phase space with $s=0$. The tightest bound is immediately obtained as $V^{-\frac{1}{3}}_{\mathrm{QSL}}=\frac{3\sqrt{3}}{4\hbar}\Delta E$. }

\begin{figure}[t]
	\centering
	\includegraphics[width=8.0cm]{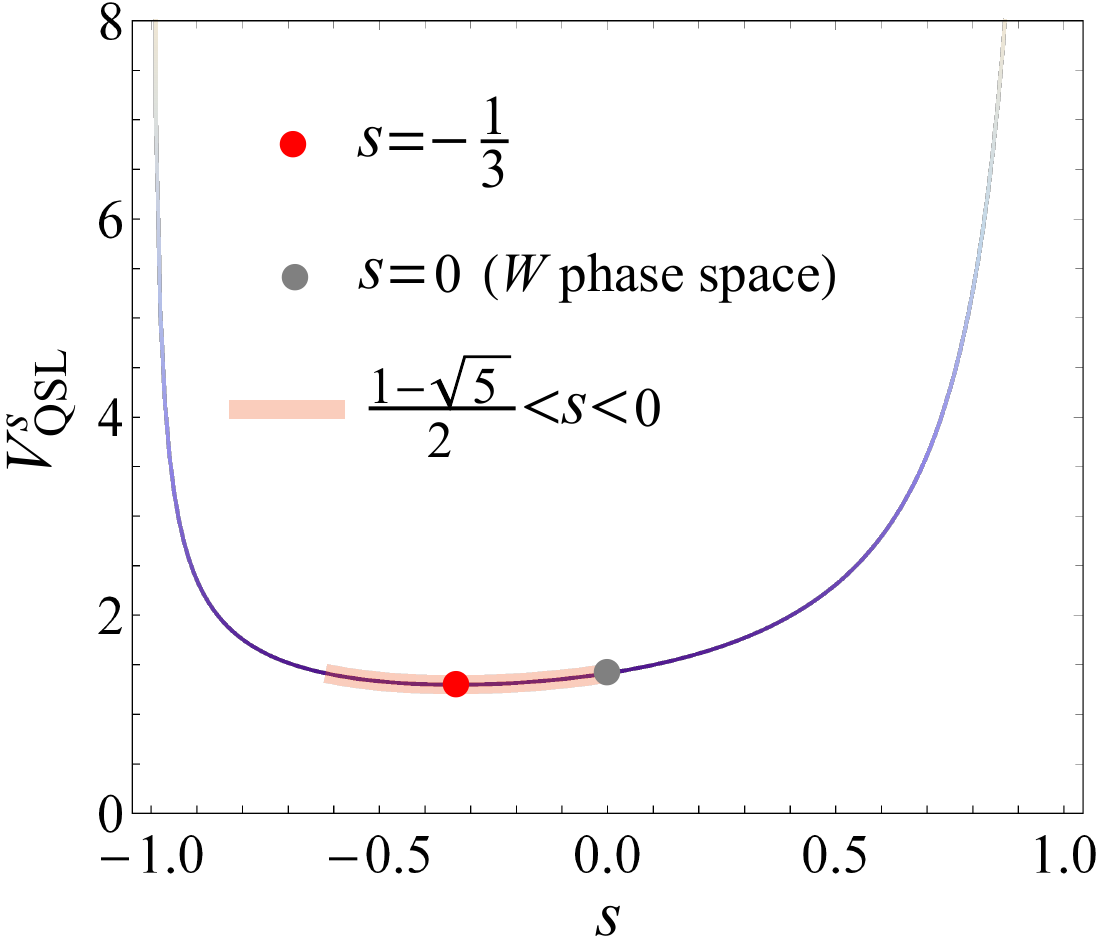}
	\caption{$V^s_{\mathrm{QSL}}$ of a single-mode free field or a one-dimensional harmonic oscillator initially prepared in state $\ket{\alpha_0}$ and evolved under the control of the Hamiltonian $H=\hbar \omega\left(a^{\dagger} a+\frac{1}{2}\right)$ ($\omega |\alpha_0|=1$). The gray dot represents $s=0$, i.e., the Wigner phase space (or the equivalent Hilbert space). {\R The orange region represents $V^s_{\mathrm{QSL}}<V^0_{\mathrm{QSL}}$} The red dot marks the lowest value of $V^s_{\mathrm{QSL}}$, implying that the tightest QSL bound does not occur in the self-dual case but in dual phase spaces with $s=-1/3$.}
	\label{fig2}
\end{figure}

In Fig. \ref{fig2}, $V^{s}_{\mathrm{QSL}}$ versus parameter $s$ is depicted given $\omega |\alpha_0|=1$. One can find that the self-dual case, i.e., $s=0$, is not the tightest one (neither are the other well-known representations, e.g., $s=-1$). Obviously, there is a representation with $s\neq 0$ that has the lowest upper bound. Indeed, by choosing an appropriate parameter of the phase space, i.e., $s=-1/3$ {\ZY as analyzed above}, we can find the corresponding tightest dual phase spaces, as marked by the red dot in Fig. \ref{fig2}.

{\ZY The above analysis implies that the $s$-parametrized phase spaces provide an approach to search for tighter QSL bounds than the one given initially in $W$ phase space or Hilbert space. Such a phase space is unlike the usual $W$, $P$, and $Q$ representations. Indeed, the rarely used phase spaces with $s\neq 0, \pm 1$ may also be helpful in specific quantum tasks, such as searching for the tightest QSL bound, as illustrated here.}

{\ZY Thus far, our example involves a continuous variable system. It is natural to ask if this universal QSL applies to finite-dimensional systems in the same framework and whether the above conclusion still holds in such cases. }

\subsection{{\ZY $N$-level quantum systems} in a continuous phase space}\label{subSecI}
Consider an {\ZY $N$-level quantum system with the following Hamiltonian given by \cite{Eberly1981PRL}}
\begin{equation}
	H=h_0 \mathbbm{1}+h_{\nu} T_{\nu}, \label{HamiltonSU(N)}
\end{equation}
where $h_0$ and $h_\nu$ are coefficients, e.g., a magnetic field. $T_{\nu} (\nu=1,2,\cdots,N^2-1)$, a set of traceless Hermitian matrices, are the generators of $\mathrm{SU}(N)$ Lie algebra \cite{Haber2021}, and the summation over repeated Greek indices is implicit.  

The SW kernel of the {\ZY $N$-level quantum system} is defined as \cite{Runeson2020,Tilma_2012} {\ZY (see Appendix \ref{SWc})}
\begin{equation}
	\Delta^s(\eta)=\frac{1}{N} \mathbbm{1}+4 r_{s} R_\nu T_\nu, \label{kernelSU(N)}
\end{equation}
where $ R_\nu:=\bra{\eta} T_\nu \ket{\eta} $, and
\begin{equation}
    r_s=\frac{1}{2}\sqrt{(N+1)^{1+s}} \label{rs}
\end{equation}
is called the $s$-parametrized spin radius, representing different phase spaces. {\ZY Unlike the example in Sec. \ref{subSec3}, there is no restriction regarding the parameter $s$ here, i.e., $s\in \mathbbm{R}$.} Following the terminology in quantum optics, we still deem $s=0$ the $W$ phase space, while $s=1$ and $s=-1$ represent the $P$ and $Q$ phase spaces, respectively \cite{Runeson2020}.

The evolved density matrix in terms of the basis of the $\mathrm{SU}(N)$ Lie algebra is represented by \cite{Eberly1981PRL}
\begin{equation}
	\rho_t=\frac{1}{N} \mathbbm{1}+b_{\nu}(t) T_\nu, \label{rhoSU(N)} 
\end{equation}
with $b_\nu (t):=2\mathrm{Tr}(\rho_t T_\nu)$, where we have employed $ \mathrm{Tr}(T_\nu T_\lambda)=\frac{1}{2}\delta_{\nu\lambda} $ \cite{Haber2021}. 

With Eqs. (\ref{HamiltonSU(N)}), (\ref{kernelSU(N)}), and (\ref{rhoSU(N)}), the time derivative of the phase space function $F^s_{\rho_t}(\eta)$ in Eq. (\ref{dynamics}) is obtained $\partial_t F^s_{\rho_t}(\eta)=\frac{2}{\hbar} r_s h_\nu b_\lambda (t) R_\xi f_{\nu \lambda \xi}$, where we use $[T_\nu,T_\lambda]=i f_{\nu\lambda\xi} T_\xi$ and $f_{\nu\lambda\xi}$ are totally antisymmetric regarding the interchange of any pair of its indices \cite{Haber2021}. Then, we have
\begin{equation}
	\chi^s_t=\left [\frac{1}{N}+\frac{2}{N+1} b^2_\nu (t) r^2_{s} \right ]^\frac{1}{2},   \label{upB1}
\end{equation}
according to $\int d \mu(\eta) R_\nu R_\lambda=\frac{1}{2(N+1)} \delta_{\nu \lambda}$ \cite{Runeson2020}. In addition, the integration in Eq. (\ref{vQSL_t}) yields
\begin{equation}
    v^s_{\mathrm{QSL}}(t)=\frac{r_s}{\hbar} \left[\frac{2}{(N+1)} h_\nu b_\lambda (t) h_\beta b_\gamma (t) f_{\nu \lambda \xi} f_{\beta \gamma \xi} \right]^{\frac{1}{2}}. \label{upB2}
\end{equation}
{\ZY With Eqs. (\ref{upB1}) and (\ref{upB2}), the QSL bound in Eq. (\ref{VQSL}) is immediately obtained.}

In addition, an interesting corollary based on Eq. (\ref{upB2}) is obtained, $v^0_{\mathrm{QSL}}(t)^2=v^{-s}_{\mathrm{QSL}}(t)v^s_{\mathrm{QSL}}(t)$, which can be proven immediately based on the definition of the $s$-parametrized spin radius in Eq. (\ref{rs}). This corollary implies that one component of the QSL bounds, $v^s_{\mathrm{QSL}}(t)$ with $s \neq 0$, could be tighter than that in the $s=0$ case at the expense of a looser bound for $-s$. Does this conclusion still hold for $V^{s}_{\mathrm{QSL}}(t)$? {\ZY However, $V^{0}_{\mathrm{QSL}}(t)^2 \neq V^{-s}_{\mathrm{QSL}}(t)V^{s}_{\mathrm{QSL}}(t)$ in general, as the other component of $V^{s}_{\mathrm{QSL}}(t)$, i.e., $\chi^s_t$, is not a constant.} Nevertheless, if we carefully choose dual phase spaces, it is still possible to obtain a QSL bound tighter than the usual bound in $W$ phase space or Hilbert space [$V^{0}_{\mathrm{QSL}}(t)$ is identical to the speed limit obtained in Hilbert space; see the proof in Appendix \ref{Appendix3}].
{\R Indeed, if the initial state is pure, we have 
\begin{equation}
    \frac{V^{0}_{\mathrm{QSL}}}{V^{s}_{\mathrm{QSL}}}=\left[ \frac{N}{(N+1)^s+N-1} \right] ^ \frac{1}{2}, \label{rationV}
\end{equation}
implying $V^{s<0}_{\mathrm{QSL}}<V^{0}_{\mathrm{QSL}}$. The maximum of Eq. (\ref{rationV}) is $\sqrt{N/(N-1)}$ when $s$ approaches $-\infty$, signifying that the upper bound of the QSL originally derived in $W$ phase space will, at most, decrease by $(1-\sqrt{1-1/N)}$ $\times$ $100\%$ in $s$-parametrized phase spaces. }

For illustration, we consider $N=2$, i.e., a qubit system. The Hamilton in Eq. (\ref{HamiltonSU(N)}) takes the form
\begin{equation}
    H=h_0 \mathbbm{1}+h_j T_j=h_0 \mathbbm{1}+\frac{1}{2}\vec{h} \cdot \vec{\sigma}, \label{Hqubit}
\end{equation}
where $ T_j=\frac{1}{2} \sigma_j (j=x,y,z)$ {\ZY and $\sigma_j$ are the usual Pauli operators}. We replace the Greek indices with the Latin ones, as usual, and thus, $ f_{jkl} $ equals $ \varepsilon_{jkl} $ (the three-dimensional Levi-Civita symbol) \cite{Haber2021}. The evolved density matrix in Eq. (\ref{rhoSU(N)}) now reads 
\begin{equation}
    \rho_t=\frac{1}{2} \mathbbm{1}+b_j (t) T_j=\frac{1}{2} \mathbbm{1}+\frac{1}{2}\vec{b}_t \cdot \vec{\sigma}. \label{rhoqubit}
\end{equation}
After some algebra, Eqs. (\ref{upB1}) and (\ref{upB2}) can be simplified to
\begin{equation}
    \chi^s_t=\left [\frac{1}{2}+\frac{2}{3} |\vec{b}_t|^2 r^2_{s} \right ]^\frac{1}{2} \label{chiqubit1}
\end{equation}
and
\begin{equation}
    v^s_{\mathrm{QSL}}(t)=\frac{1}{\hbar}\sqrt{\frac{2}{3}} r_s |\vec{h} \times \vec{b}_t|, \label{vqubit1}
\end{equation}
where $ \varepsilon_{jkp} \varepsilon_{lmp}=\delta_{jl}\delta_{km}-\delta_{jm}\delta_{kl} $ is employed. {\ZY Thus, 
\begin{eqnarray}
    V_{\mathrm{QSL}}^s(t)= & \frac{1}{2 \hbar} \min \left\{\sqrt{3^s+\left|\vec{b}_0\right|^2}\left|\vec{h} \times \vec{b}_t\right|,\right. \notag\\
& \left.\sqrt{3^s+\left|\vec{b}_t\right|^2}\left|\vec{h} \times \vec{b}_0\right|\right\}.\label{VQSLqubit1}
\end{eqnarray}
Former research found that the QSL bound derived based on the metric of relative purity will be proportional to the standard deviation of energy \cite{Adolfo2013PRL}. When the initial state is pure, according to Eqs. (\ref{Hqubit}) and (\ref{rhoqubit}), we have $\Delta E=\frac{1}{2} |\vec{h} \times \vec{b}_t|=\frac{1}{2} |\vec{h} \times \vec{b}_0|$ and $\left|\vec{b}_t\right|^2=\left|\vec{b}_0\right|^2=1$. Thus, the $s$-parametrized QSL bound of qubit systems in Eq.(\ref{VQSLqubit1}) is rewritten as 
\begin{equation}
    V^s_{\mathrm{QSL}}=\sqrt{1+3^s} \frac{\Delta E}{\hbar}, \label{vqubitBound}
\end{equation}
which is a Mandelstam-Tamm type bound. When $s<0$, the QSL bounds will be tighter than the one in Wigner phase space, i.e., $V^{s<0}_{\mathrm{QSL}}<V^0_{\mathrm{QSL}}$ and the tightest bound achieves when $s\rightarrow -\infty$. Thus, the quasiprobability distribution phase spaces can be employed to search tighter QSL bound, which agrees with the example illustrated in Sec \ref{subSec3}. }

%Note that we are not aiming to compare the values of the QSL derived in the paper, and the original MT bound ($V^{\mathrm {MT}}_{\mathrm{QSL}}=\frac{2 \Delta E}{\pi \hbar}$) because they are under different metrics.  Comparing the bounds under the same metric is more meaningful, as in the main text for QSLs in different phase spaces under the same relative purity metric.

\begin{figure}[t]
	\centering
	\includegraphics[width=8.0cm]{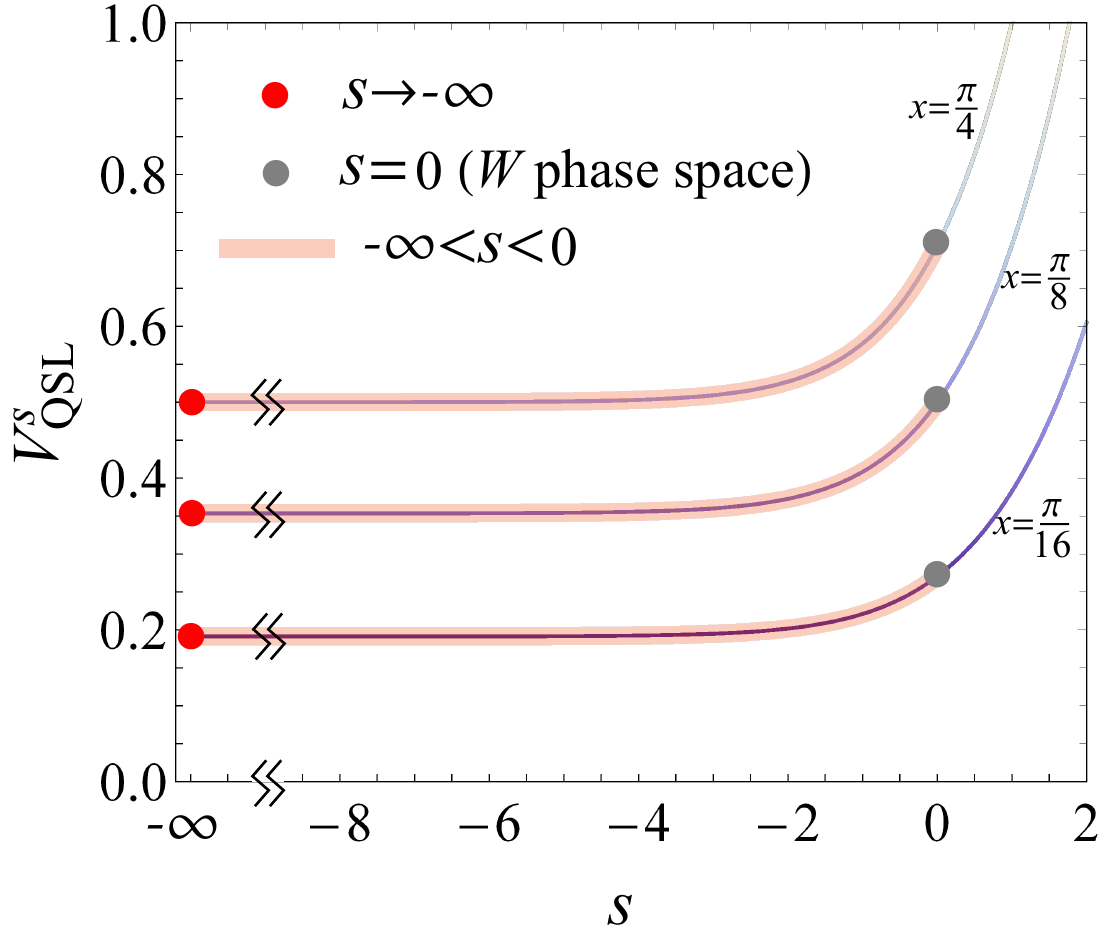}
	\caption{$V^s_{\mathrm{QSL}}$ of a qubit initially prepared in state $\ket{\psi}=\cos(x)\ket{0}+\sin(x) \ket{1}$ and evolved under the control of the Hamiltonian $H=\frac{\hbar}{2}\sigma_z$. The blue curves from bottom to top represent {\R $V^s_{\mathrm{QSL}}$ with $x=\frac{\pi}{16}$, $\frac{\pi}{8}$, and $\frac{\pi}{4}$, respectively. Within the orange regions, $V^s_{\mathrm{QSL}}<V^0_{\mathrm{QSL}}$. The red and gray dots denote $V^{-\infty}_{\mathrm{QSL}}=\frac{1}{2}|\sin(2x)|$ and $V^{0}_{\mathrm{QSL}}=\frac{1}{\sqrt{2}}|\sin(2x)|$.} }
	\label{fig1}
\end{figure}
%The gray circle represents the QSL in Hilbert space. The bound for $s=-1$ is tighter than the one in Hilbert space, which is at the expense of a looser bound for $s=1$.

For illustration, in Fig. \ref{fig1}, we consider an initial state $\ket{\psi}=\cos(x)\ket{0}+\sin(x) \ket{1}$ under the control of the Hamiltonian $H=\frac{\hbar}{2}\sigma_z$. Thus, the QSL reads 
\begin{equation}
	V^{s}_{\mathrm{QSL}}=\frac{1}{2} \sqrt{1+ 3^s} \left\vert  \sin(2x)\right\vert.
\end{equation}
Obviously, the QSL bound in phase space $s=-1$ is tighter than that with $s=0$.  Thus, the tightest bound occurs when $s$ is approaching $-\infty$, and the bound reads $V^{-\infty}_{\mathrm{QSL}}=\frac{1}{2}|\sin(2x)|$ (see the red dots in Fig. \ref{fig1}). 

Unlike the continuous phase space representation, another equivalent approach is to employ the discrete phase space in an $N \times N$ lattice form {\R (see Appendix \ref{AppendixE}). }

The above examples illustrate that both for continuous variable and finite-dimensional discrete quantum systems, we may find tighter QSL bound in specific phase spaces than in $W$ phase space or Hilbert space {\ZY in terms of the  quasiprobability distribution phase spaces}. Importantly, this phenomenon is detectable in experiments.

\section{Experimental consideration}\label{EP}
Given an initial state $\rho_0$ as well as the Hamiltonian $H$, it would be more convenient to verify $\chi_t^{-s}v^s_{\mathrm{QSL}}(0)$. We need only to measure the phase space function of an evolved state, i.e., $F^{-s}_{\rho_t} (\eta)=\mathrm{Tr}[\rho_t \Delta^{-s} (\eta)]$. The key procedure is constructing the SW kernel $\Delta^s (\eta)$, which can be decomposed to a combination of spin-parity operators $\Pi^s$ and displacement-rotation operators $U$ \cite{Tilma2016PRL}. For example, the SW kernel of a qubit can be rewritten as $\Delta^s(\eta)=U(\theta,\phi,\Phi)\Pi^s U^{\dagger}(\theta,\phi,\Phi)$, where $U(\theta,\phi,\Phi)=\mathrm{exp}(-\frac{i\sigma_{z}}{2}\phi)\mathrm{exp}(-\frac{i\sigma_{y}}{2}\theta) \mathrm{exp}(-\frac{i\sigma_{z}}{2}\Phi)$ and $\Pi^s=\frac{1}{2}\mathbbm{1}+r_{s}\sigma_{z}$. A similar scheme for verifying the Wigner functions has been realized in experiments with a cesium atom \cite{ZhangTiancai2018PRA} and an electron spin of a nitrogen-vacancy center in diamond \cite{XuNanyang2019APL}. 

For continuous variable systems, the SW kernel can also be constructed as a displaced parity operator, i.e., $\Delta^s (\eta)=D(\eta)\Pi^s D^{\dagger}(\eta)$ \cite{Rundle2021,Davidovich1997PRL}. Thus, the phase space function $F^{-s}_{\rho_t} (\eta)$ is just the inverse Fourier transform of the characteristic function $G^{-s}_{\rho_t} (\eta)=\mathrm{Tr}[\rho_t D (\eta)e^{-\frac{1}{2} s|\eta|^2}]$. It is feasible to measure this characteristic function in experiments by means of single-qubit interferometry; see, e.g., Refs. \cite{Tufarelli2012PRA,Zhenyu2019PRL,Zhenyu2022PRR}.

\section{Conclusion and outlook}
To conclude, we established a universal QSL bound based on the SW correspondence. This speed limit bound allows us to investigate the dynamics of {\ZY both continuous variable and finite-dimensional discrete} quantum systems in arbitrary phase spaces in the same framework as long as the SW kernel is given. Remarkably, by choosing specific dual phase spaces, the QSL bound can be tighter than that in the $W$ phase space or Hilbert space, which is demonstrated for several typical continuous variable and finite-dimensional discrete quantum systems. This conclusion allows us to search for tighter QSL bounds {\ZY in terms of specific quasiprobability distribution phase spaces and provides a real example that rarely used phase spaces with $s \neq 0$, $\pm 1$ may be superior to the commonly used $W$, $P$, and $Q$ phase spaces in certain quantum tasks}. 

Our results suggest several areas for further investigation. First, it would be of great interest to investigate the QSL with $\hbar$ deformation in discrete quantum systems in addition to continuous variable systems. If one takes $\hbar$ as a quantum parameter, $\hbar$=0 gives the classical Poisson bracket in the standard Moyal product. Comparing the dynamics for different $\hbar$ values for QSL with general SW formalism is fascinating and may provide a thorough understanding of the speed limit on the border between the macroscopic and microscopic world. Second, whether we could utilize the tighter QSL bound in specific dual phase spaces superior to the self-dual case (Hilbert space) and design schemes for operating logical gates in such phase spaces would be of interest.

\section{Acknowledgements}
The authors wish to thank Adolfo del Campo for enlightening discussions and helpful comments. They also wish to thank Fernando J. G$\acute{\mathrm{o}}$mez-Ruiz for the helpful feedback. This work was supported by the National Natural Science Foundation of China under Grant No. 12074280.

\appendix\label{app}
\begin{widetext}

\section{Stratonovich-Weyl correspondence} \label{SWc}
This section briefly summarizes the basic idea of the Stratonovich–Weyl (SW) correspondence that will be employed in the main text. The operator $A$ in Hilbert space and the corresponding phase space function $F^s_{A}(\eta)$ are linked by a kernel $\Delta^s(\eta)$, where $\eta$ is a phase space point in any phase space labeled by the index $s$. There are five criteria for the kernel that ensure physically motivating properties. Such properties, known as the SW correspondence \cite{Rundle2021}, are listed below:

SW-1: Linearity: $A \leftrightarrow F^s_{A}(\eta)$ is a linear bijective mapping, i.e., 
\begin{equation}
    F^s_{A}(\eta)=\mathrm{Tr}[A \Delta^s(\eta)],  \label{SW1-1}
\end{equation}
and
\begin{equation}
    A=\int d\mu(\eta) F^s_{A}(\eta) \Delta^{-s}(\eta),  \label{SW1-2}
\end{equation}
where $d\mu(\eta)$ is the invariant integration measure.

SW-2: Reality: The kernel $\Delta^s(\eta)$ is Hermitian. Thus, the phase space function is real if the operator $A$ is Hermitian, i.e.,
\begin{equation}
    F^{s}_{A}(\eta)^*=\mathrm{Tr}[A^\dagger \Delta^{s} (\eta)^\dagger]=\mathrm{Tr}[A \Delta^s(\eta)]=F^s_{A}(\eta). \label{SW2}
\end{equation}

SW-3: Standardization: $F^s_{A}(\eta)$ is standardized so that 
\begin{equation}
    \int d\mu(\eta) F^s_{A}(\eta)=\mathrm{Tr}(A), \label{SW3-1}
\end{equation}
and 
\begin{equation}
    \int d\mu(\eta) \Delta^s(\eta)=\mathbbm{1}. \label{SW3-2}
\end{equation}

SW-4: Covariance: If the operator is invariant under the unitary operations $U$, then so is the phase space function, i.e.,
\begin{equation}
    F_{UAU^\dagger}^s(\eta)=\mathrm{Tr}[UAU^\dagger \Delta^s(\eta)]=\mathrm{Tr}[A V \Delta^s(\eta)V^\dagger]=\mathrm{Tr}[A \Delta^s(\eta')]=F^s_A(\eta'), \label{SW4}
\end{equation}
where $\Delta^s(\eta')=V \Delta^s(\eta)V^\dagger$ and $V=U^\dagger$. 

SW-5: Tracing:
\begin{equation}
    \int d\mu(\eta) F^s_A (\eta)F_B^{-s}(\eta)=\mathrm{Tr}(AB). \label{SW5}
\end{equation}

Note that if the phase space is discrete, the integrations in above SW-1$ \sim $5 will be replaced by the summation over the phase points.

{\ZY For illustration, we prove that Eq. (\ref{kernelSU(N)}), i.e., 
\begin{equation}
      \Delta^s(\eta)=\frac{1}{N} \mathbbm{1}+4 r_{s} R_\nu T_\nu, \label{kernelN}
\end{equation}
in the main text satisfies the above five criteria for an SW kernel. Here $r_s=\frac{1}{2}\sqrt{(N+1)^{1+s}} (s\in \mathbbm{R})$, $ R_\nu:=\bra{\eta} T_\nu \ket{\eta} $, and $T_\nu$ ($\nu=1,2,\cdots, N^2-1$) are the basis for the set of traceless hermitian $N \times N$ matrices. 

\textit{Proof.} (i) According to Eq. (\ref{SW1-1}), the phase space function of operator A is given by
\begin{equation}
        F^s_{A}(\eta)=\frac{1}{N} \mathrm{Tr}A+4r_s R_\nu \mathrm{Tr}(AT_\nu), \label{proofF}
\end{equation}   
with which we have
\begin{equation}
        \int d\mu(\eta) F^s_{A}(\eta) \Delta^{-s}(\eta)=\frac{\mathrm{Tr} A}{N} \mathbbm{1}+2\mathrm{Tr}(AT_\nu)T_\nu=A, \label{proofi}
\end{equation}
where we have employed three identities \cite{Brif1999,Runeson2020}, i.e., 
\begin{equation}
    \int d\mu(\eta)=N, \label{E1}
\end{equation}
\begin{equation}
      \int R_\nu d\mu(\eta)=0, \label{E2}
\end{equation}
and
\begin{equation}
    \int R_\nu R_\lambda d\mu(\eta)=\frac{1}{2(N+1)} \delta_{\nu \lambda}, \label{E3}
\end{equation}
in deriving the first equality in Eq. (\ref{proofi}). The second equality in Eq. (\ref{proofi}) is based on the fact that any $N \times N$ complex matrix can be expressed as a complex linear combination of the identity matrix and $N^2-1$ generators $T_\nu$ of the SU($N$) Lie algebra [see, e.g., Eq. (9) in Ref. \cite{Haber2021}]. 

(ii) The kernel Eq. (\ref{kernelN}) is obviously Hermitian.

(iii) The integration of the phase space function $F^s_{A}(\eta)$ over all space is given by
\begin{equation}
        \int d\mu(\eta) F^s_{A}(\eta)=\int d\mu(\eta) \left[ \frac{1}{N} \mathrm{Tr}A+4r_s R_\nu \mathrm{Tr}(AT_\nu)\right] = \mathrm{Tr}(A).
\end{equation} 
In addition, 
\begin{equation}
        \int d\mu(\eta)\Delta^s(\eta)=\int d\mu(\eta) \left( \frac{1}{N} \mathbbm{1}+4 r_{s} R_\nu T_\nu \right)=\mathbbm{1}. 
\end{equation}

(iv) According to SW-4, we define
\begin{equation}
       V \Delta^s(\eta)V^\dagger=\frac{1}{N}\mathbbm{1}+4r_sR_\nu VT_\nu V^\dagger=\frac{1}{N}\mathbbm{1}+4r_sR'_\nu T'_\nu =:\Delta^s(\eta'),
\end{equation}
where $T'_\nu=VT_\nu V^\dagger$, $R'_\nu=\bra{\eta'} T'_\nu \ket{\eta'}$ with $\ket{\eta'}=V\ket{\eta}$. Thus, if the operator is invariant under the unitary operations $V=U^\dagger$, then so is the phase space function.

(v) According to Eq. (\ref{proofi}), we have
\begin{equation}
        \mathrm{Tr}(AB)=\frac{1}{N}\mathrm{Tr}(A)\mathrm{Tr}(B)+2\mathrm{Tr}(AT_\nu)\mathrm{Tr}(BT_\nu).
\end{equation}
On the other hand, 
\begin{align}
      \int d\mu(\eta) F^s_A (\eta)F_B^{-s}(\eta)=&\int d\mu(\eta) \left[ \frac{1}{N} \mathrm{Tr}(A)+4r_s R_\nu \mathrm{Tr}(AT_\nu)\right] \left[ \frac{1}{N} \mathrm{Tr}(B)+4r_{-s} R_\lambda \mathrm{Tr}(BT_\lambda)\right] \notag\\=&\frac{1}{N}\mathrm{Tr}(A)\mathrm{Tr}(B)+2\mathrm{Tr}(AT_\nu)\mathrm{Tr}(BT_\nu)
\end{align}
is obtained by means of Eqs. (\ref{proofF}),(\ref{E1}$\sim$ \ref{E3}). Therefore, SW-5 is also satisfied.

To this end, we conclude that Eq. (\ref{kernelSU(N)}) used in the main text is an SW kernel.
}

\section{Derivation of the second line in Eq. (6)} \label{appB}
The changing rate of the relative purity is given by
{\ZY
\begin{equation}
    \dot{P}_{t}(\rho _{0},\rho _{t}) =\mathrm{Tr}(\rho_0 \dot{\rho}_t)=\frac{1}{i\hbar} \mathrm{Tr}(\rho_0 [H,\rho_t]).
\end{equation}
According to Eqs. (\ref{FA}) and (\ref{TrAB}), we have
\begin{equation}
    \dot{P}_{t}(\rho _{0},\rho _{t}) =\frac{1}{i\hbar} \int d\mu (\eta )F^{-s}_{\rho _{0}}(\eta)\mathrm{Tr}\left[ [H,\rho_t] \Delta^s(\eta)\right]. \label{Ptt}
\end{equation}
Then, substituting the phase space form of $H=\int d\mu (\eta' )F^{s'}_H (\eta') \Delta^{-s'}(\eta') $ and $\rho=\int d\mu (\eta'' )F^{s''}_{\rho_t} (\eta'') \Delta^{-s''}(\eta'')$ in Eq. (\ref{Ptt}), we obtain
\begin{equation}
    \dot{P}_{t}(\rho _{0},\rho _{t}) =\frac{1}{i\hbar} \int d\mu (\eta )F^{-s}_{\rho _{0}}(\eta) ( F^s_{H}\star F^s_{\rho_t}-F^s_{\rho_t}\star F^s_{H})(\eta)=\int d\mu (\eta ) F^{-s}_{\rho _{0}}(\eta)\left\{ \left\{F^s_{H},F^s_{\rho _{t}}\right\} \right\}(\eta), \label{Pttt}
\end{equation}
where we have employed the generalized star product [Eq. (\ref{star product})] and Moyal bracket [Eq. (\ref{Moyal bracket})] defined in the main text. Then, by means of $\mathrm{Tr}(\rho_0 [H,\rho_t])=\mathrm{Tr}(\rho_t [\rho_0,H])$ and similar procedures performed in above Eqs. (\ref{Ptt}) and (\ref{Pttt}), we have 
\begin{equation}
    \dot{P}_{t}(\rho _{0},\rho _{t}) =\int d\mu (\eta ) F^{-s}_{\rho _{t}}(\eta)\left\{ \left\{F^s_{\rho _{0}},F^s_{H}\right\} \right\}(\eta),
\end{equation}
which is the second line in Eq. (\ref{puritychange}) in the main text.}

\section{QSLs of a single-mode free field or a one-dimensional harmonic oscillator in s-parametrized phase spaces}\label{app4}
We consider a one-dimensional harmonic oscillator or a single-mode free field prepared in a coherent state $\ket{\alpha_0}$ and evolved under the control of Hamiltonian $H=\hbar \omega\left(a^{\dagger} a+\frac{1}{2}\right)$. So the evolved density operator can be written as $\rho_{t}=\ket{\alpha}\bra{\alpha}$, where we have used the definition of $\alpha:=\alpha_0 e^{-i\omega t}$ \cite{knight2004}. Then the SW symbol of the coherent state can be derived in terms of the SW kernel 
\begin{equation}
	F^s_{\rho_t}(\eta)=\mathrm{Tr}[\rho_{t}\Delta^s(\eta)]=\mathrm{Tr}[\ket{\alpha}\bra{\alpha}D(\eta)\Pi^s D^{\dagger}(\eta)]=\kappa \exp \left( -\kappa |\alpha-\eta|^2 \right) ,
\end{equation}
where $D(\eta)$ is a displaced operator, $\Pi^s:=\frac{2}{1-s}\left(\frac{s+1}{s-1}\right)^{a^{\dagger} a}$  $(-1\leqslant s<1)$ is a $s$-parametrized parity operator \cite{BookKlimov}, and $\kappa:=2/(1-s)$. {\ZY Note that the SW kernel written in above equation is equivalent to the one in Eq. (\ref{DeltaC}).}

Then, the changing rate of $F^s_{\rho_t}(\eta)$ is obtained
\begin{equation}
    \frac{\partial F^s_{\rho_t}(\eta)}{\partial t}=i \omega \kappa^2 \left(\alpha^* \eta-\alpha \eta^*\right) \exp \left(-\kappa|\alpha-\eta|^2\right).
\end{equation}

To this end, we are ready to calculate the two components of $V^s_{\mathrm{QSL}}(t)$: 
\begin{eqnarray}
    (\chi^s_t)^2&=&\frac{\kappa^2}{\pi} \int \exp \left( -2 \kappa |\alpha-\eta|^2\right) d^2 \eta \notag\\
   &=&\frac{\kappa^2}{\pi} \exp(-2\kappa|\alpha|^2) \int \exp(2\kappa \alpha^* \eta) \exp(2\kappa \alpha \eta^*-2\kappa |\eta|^2)d^2\eta \notag\\
   &=&\frac{\kappa}{2}=\frac{1}{1-s}, \label{alphasM}
\end{eqnarray}
   
and
\begin{eqnarray}
	v^s_{\mathrm{QSL}}(t)^2&=&\frac{\omega^2 \kappa^4}{\pi} \int  \left[2|\alpha|^2|\eta|^2-\alpha^2\left(\eta^*\right)^2-\left(\alpha^*\right)^2 \eta^2\right] \exp \left(-2\kappa |\alpha-\eta|^2\right) d^2 \eta \notag\\
	&=&\frac{\omega^2 \kappa^4}{\pi} \exp(-2\kappa|\alpha|^2)  \int  \left[2|\alpha|^2|\eta|^2-\alpha^2\left(\eta^*\right)^2-\left(\alpha^*\right)^2 \eta^2\right]  \exp(2\kappa \alpha^* \eta) \exp(2\kappa \alpha \eta^*-2\kappa |\eta|^2)d^2\eta  \notag\\
	&=&\omega^2 \kappa^4 \exp(-2\kappa|\alpha|^2)  \left[\frac{|\alpha|^2 \exp(2\kappa |\alpha|^2)}{2 \kappa^2}+\frac{|\alpha|^4}{\kappa}\exp(2\kappa|\alpha|^2)-\frac{|\alpha|^4}{\kappa}\exp(2\kappa|\alpha|^2)\right] \notag\\
	&=&\frac{\kappa^2 \omega^2  |\alpha|^2 }{2 }=\frac{2}{(1-s)^2}\omega^2  |\alpha_0|^2,\label{coheretstate}
\end{eqnarray}
where we employ the formula 
\begin{equation}
    \frac{1}{\pi}\int|x|^{2m}f(x)\exp(yx^*-z|x|^2)d^2x=\frac{1}{z^{m+1}}\frac{d^m\left[\left(\frac{y}{z}\right)^mf(\frac{y}{z})\right]}{d\left(\frac{y}{z}\right)^m}, \label{F1}
\end{equation}
in the derivation of above Eqs. (\ref{alphasM}) and (\ref{coheretstate}).
Here, we summarize the basic idea for the proof of Eq. (\ref{F1}). Note that the following $x,y,z$ parameters are all complex numbers.
\begin{equation}
	\frac{1}{\pi}\int|x|^{2m}f(x)\exp(yx^*-z|x|^2)d^2x=\displaystyle\sum_n c_nK_n,
\end{equation}
where
\begin{equation}
    K_n=\frac{1}{\pi}\int|x|^{2m}x^n\frac{(x^*)^n y^n}{n!}\exp(-z|x|^2)d^2x=\frac{y^n}{\pi n!}\int|x|^{2(n+m)}\exp(-z|x|^2)d^2x=\frac{y^n(n+m)!}{z^{n+m+1}n!}.
\end{equation}
Then we have
\begin{equation}
    \frac{1}{\pi}\int|x|^{2m}f(x)\exp(yx^*-z|x|^2)d^2x=\displaystyle\sum_n c_n\frac{y^n(n+m)!}{z^{n+m+1}n!}=\frac{1}{z^{m+1}}\frac{d^m\left[\left(\frac{y}{z}\right)^mf(\frac{y}{z})\right]}{d\left(\frac{y}{z}\right)^m},
\end{equation}
which ends our proof.

With Eqs.(\ref{alphasM}) and (\ref{coheretstate}), we obtain 
\begin{equation}
    V^{s}_{\mathrm{QSL}}=\chi^{-s}v^s_{\mathrm{QSL}}=\frac{\sqrt{2}\omega |\alpha_0|}{(1-s)\sqrt{1+s}},
\end{equation}
in the main text [see Eq. (\ref{VQSL3})].

\section{QSLs of {\ZY $N$-level quantum systems} in Hilbert space}\label{Appendix3}
The QSL bound of $N$-level quantum systems is given by 
\begin{equation}
    \left\vert \dot{P}_{t}(\rho _{0},\rho _{t})\right\vert \leq V_\mathrm{QSL}(t)=\sqrt{\mathrm{Tr}\left( \rho_0^2 \right) \mathrm{Tr}\left( |\dot{\rho}_t|^2 \right) }.
    \label{boundPtH}
\end{equation}
According to the von Neumann equation and Eqs. (\ref{HamiltonSU(N)}) and (\ref{rhoSU(N)}) in the main text, we have $\dot{\rho}_t = \frac{1}{\hbar}h_\nu b_\lambda(t) f_{\nu\lambda\xi} T_\xi$. Thus
\begin{equation}
    V_\mathrm{QSL}(t)=\frac{1}{2\hbar}\sqrt{\left(\frac{2}{N}+b^2_\iota \right) h_\nu b_\lambda(t) h_\beta b_\gamma(t) f_{\nu\lambda\xi} f_{\beta\gamma\xi}},
\end{equation}
which equals to the QSL bound in $W$ phase space ($s=0$) in the main text [Eqs. (\ref{upB1}) and (\ref{upB2})].

\section{Qubits in a discrete phase space} \label{AppendixE}
The discrete phase space function represents a quantum state that assigns a number to each lattice point \cite{Vourdas2004}. We need only to replace the generalized Moyal bracket and the star product with the corresponding discrete ones.

For illustration, we still consider the qubit case ($N=2$). We can label the phase space points $\eta$ with
coordinates $(a_1,a_2)$ ($a_{1(2)}=0,1$). The corresponding SW kernel is \cite{Wootters1987}
\begin{equation}
    \Delta^s(\eta)=\frac{1}{2} \mathbbm{1}+\frac{\sqrt{3}}{4 r_s} \left[(-1)^{a_{1}} {\sigma}_{z}+(-1)^{a_{2}} {\sigma}_{x}+(-1)^{a_{1}+a_{2}} {\sigma}_{y}\right],
\end{equation}
which obviously satisfies the SW correspondence, with $\int d\mu(\eta)$ replaced by $\frac{1}{2} \sum_\eta$. Here, $r_s=\frac{1}{2} \sqrt{3^{1+s}}$, with $s=-1, 0, 1$ representing the $Q$, $W$, and $P$ representations, respectively.
By replacing the generalized star product in Eq. (\ref{star product}) with $(F^s_A \star F^s_B)(\eta)=\frac{\Gamma_{\eta \beta \gamma}}{2} F^s_A(\beta) F^s_B(\gamma)$ \cite{Wootters1987}, we can obtain the QSL bound. Here, $\Gamma_{\eta \beta \gamma}=\delta_{\eta \beta}+\delta_{\eta \gamma}+\delta_{\beta \gamma}-\frac{1}{2}-i \sum_\kappa\varepsilon_{\eta \beta \gamma \kappa}$ is the three-point structure function, and $\varepsilon$ is antisymmetric in all of its indices. We then immediately have $\chi^s_t=\left [\frac{1}{2}+\frac{2}{3} |\vec{b}_t|^2 r^2_{s} \right ]^\frac{1}{2}$ and $v^s_{\mathrm{QSL}}(t)=\frac{1}{\hbar}\sqrt{\frac{2}{3}} r_s |\vec{h} \times \vec{b}_t|$, which are in agreement with $N=2$ in Sec. \ref{subSecI}.

\end{widetext}

\bibliography{references_QSL}

%apsrev4-2.bst 2019-01-14 (MD) hand-edited version of apsrev4-1.bst
%Control: key (0)
%Control: author (8) initials jnrlst
%Control: editor formatted (1) identically to author
%Control: production of article title (-1) disabled
%Control: page (0) single
%Control: year (1) truncated
%Control: production of eprint (0) enabled
\begin{thebibliography}{106}%
\makeatletter
\providecommand \@ifxundefined [1]{%
 \@ifx{#1\undefined}
}%
\providecommand \@ifnum [1]{%
 \ifnum #1\expandafter \@firstoftwo
 \else \expandafter \@secondoftwo
 \fi
}%
\providecommand \@ifx [1]{%
 \ifx #1\expandafter \@firstoftwo
 \else \expandafter \@secondoftwo
 \fi
}%
\providecommand \natexlab [1]{#1}%
\providecommand \enquote  [1]{``#1''}%
\providecommand \bibnamefont  [1]{#1}%
\providecommand \bibfnamefont [1]{#1}%
\providecommand \citenamefont [1]{#1}%
\providecommand \href@noop [0]{\@secondoftwo}%
\providecommand \href [0]{\begingroup \@sanitize@url \@href}%
\providecommand \@href[1]{\@@startlink{#1}\@@href}%
\providecommand \@@href[1]{\endgroup#1\@@endlink}%
\providecommand \@sanitize@url [0]{\catcode `\\12\catcode `\$12\catcode
  `\&12\catcode `\#12\catcode `\^12\catcode `\_12\catcode `\%12\relax}%
\providecommand \@@startlink[1]{}%
\providecommand \@@endlink[0]{}%
\providecommand \url  [0]{\begingroup\@sanitize@url \@url }%
\providecommand \@url [1]{\endgroup\@href {#1}{\urlprefix }}%
\providecommand \urlprefix  [0]{URL }%
\providecommand \Eprint [0]{\href }%
\providecommand \doibase [0]{https://doi.org/}%
\providecommand \selectlanguage [0]{\@gobble}%
\providecommand \bibinfo  [0]{\@secondoftwo}%
\providecommand \bibfield  [0]{\@secondoftwo}%
\providecommand \translation [1]{[#1]}%
\providecommand \BibitemOpen [0]{}%
\providecommand \bibitemStop [0]{}%
\providecommand \bibitemNoStop [0]{.\EOS\space}%
\providecommand \EOS [0]{\spacefactor3000\relax}%
\providecommand \BibitemShut  [1]{\csname bibitem#1\endcsname}%
\let\auto@bib@innerbib\@empty
%</preamble>
\bibitem [{\citenamefont {Pfeifer}\ and\ \citenamefont
  {Fr\"ohlich}(1995)}]{Pfeifer2005RMP}%
  \BibitemOpen
  \bibfield  {author} {\bibinfo {author} {\bibfnamefont {P.}~\bibnamefont
  {Pfeifer}}\ and\ \bibinfo {author} {\bibfnamefont {J.}~\bibnamefont
  {Fr\"ohlich}},\ }\href {https://doi.org/10.1103/RevModPhys.67.759} {\bibfield
   {journal} {\bibinfo  {journal} {Rev. Mod. Phys.}\ }\textbf {\bibinfo
  {volume} {67}},\ \bibinfo {pages} {759} (\bibinfo {year} {1995})}\BibitemShut
  {NoStop}%
\bibitem [{\citenamefont {Busch}(2008)}]{BookBusch}%
  \BibitemOpen
  \bibfield  {author} {\bibinfo {author} {\bibfnamefont {P.}~\bibnamefont
  {Busch}},\ }\href@noop {} {\emph {\bibinfo {title} {The time-energy
  uncertainty relation, in Time in Quantum Mechanics, edited by J. Muga, R. S.
  Mayato, and Í. Egusquiza}}}\ (\bibinfo  {publisher} {Springer},\ \bibinfo
  {address} {Berlin, Heidelberg},\ \bibinfo {year} {2008})\BibitemShut
  {NoStop}%
\bibitem [{\citenamefont {Deffner}\ and\ \citenamefont
  {Campbell}(2017)}]{Deffner2017Review}%
  \BibitemOpen
  \bibfield  {author} {\bibinfo {author} {\bibfnamefont {S.}~\bibnamefont
  {Deffner}}\ and\ \bibinfo {author} {\bibfnamefont {S.}~\bibnamefont
  {Campbell}},\ }\href {https://doi.org/10.1088/1751-8121/aa86c6} {\bibfield
  {journal} {\bibinfo  {journal} {Journal of Physics A: Mathematical and
  Theoretical}\ }\textbf {\bibinfo {volume} {50}},\ \bibinfo {pages} {453001}
  (\bibinfo {year} {2017})}\BibitemShut {NoStop}%
\bibitem [{\citenamefont {Mandelstam}\ and\ \citenamefont
  {Tamm}(1945)}]{MTbound1945}%
  \BibitemOpen
  \bibfield  {author} {\bibinfo {author} {\bibfnamefont {L.}~\bibnamefont
  {Mandelstam}}\ and\ \bibinfo {author} {\bibfnamefont {I.}~\bibnamefont
  {Tamm}},\ }\href {https://doi.org/10.1007/978-3-642-74626-0_8} {\bibfield
  {journal} {\bibinfo  {journal} {J. Phys. USSR}\ }\textbf {\bibinfo {volume}
  {9}},\ \bibinfo {pages} {249} (\bibinfo {year} {1945})}\BibitemShut {NoStop}%
\bibitem [{\citenamefont {Bhattacharyya}(1983)}]{Bhattacharyya_1983}%
  \BibitemOpen
  \bibfield  {author} {\bibinfo {author} {\bibfnamefont {K.}~\bibnamefont
  {Bhattacharyya}},\ }\href {https://doi.org/10.1088/0305-4470/16/13/021}
  {\bibfield  {journal} {\bibinfo  {journal} {Journal of Physics A:
  Mathematical and General}\ }\textbf {\bibinfo {volume} {16}},\ \bibinfo
  {pages} {2993} (\bibinfo {year} {1983})}\BibitemShut {NoStop}%
\bibitem [{\citenamefont {Anandan}\ and\ \citenamefont
  {Aharonov}(1990)}]{AA1990PRL}%
  \BibitemOpen
  \bibfield  {author} {\bibinfo {author} {\bibfnamefont {J.}~\bibnamefont
  {Anandan}}\ and\ \bibinfo {author} {\bibfnamefont {Y.}~\bibnamefont
  {Aharonov}},\ }\href {https://doi.org/10.1103/PhysRevLett.65.1697} {\bibfield
   {journal} {\bibinfo  {journal} {Phys. Rev. Lett.}\ }\textbf {\bibinfo
  {volume} {65}},\ \bibinfo {pages} {1697} (\bibinfo {year}
  {1990})}\BibitemShut {NoStop}%
\bibitem [{\citenamefont {Margolus}\ and\ \citenamefont
  {Levitin}(1998)}]{TLbound1998}%
  \BibitemOpen
  \bibfield  {author} {\bibinfo {author} {\bibfnamefont {N.}~\bibnamefont
  {Margolus}}\ and\ \bibinfo {author} {\bibfnamefont {L.~B.}\ \bibnamefont
  {Levitin}},\ }\href
  {https://doi.org/https://doi.org/10.1016/S0167-2789(98)00054-2} {\bibfield
  {journal} {\bibinfo  {journal} {Physica D: Nonlinear Phenomena}\ }\textbf
  {\bibinfo {volume} {120}},\ \bibinfo {pages} {188} (\bibinfo {year}
  {1998})}\BibitemShut {NoStop}%
\bibitem [{\citenamefont {Luo}\ and\ \citenamefont {Zhang}(2005)}]{Luo2005}%
  \BibitemOpen
  \bibfield  {author} {\bibinfo {author} {\bibfnamefont {S.}~\bibnamefont
  {Luo}}\ and\ \bibinfo {author} {\bibfnamefont {Z.}~\bibnamefont {Zhang}},\
  }\href {https://doi.org/10.1007/s11005-004-5095-4} {\bibfield  {journal}
  {\bibinfo  {journal} {Letters in Mathematical Physics}\ }\textbf {\bibinfo
  {volume} {71}},\ \bibinfo {pages} {1} (\bibinfo {year} {2005})}\BibitemShut
  {NoStop}%
\bibitem [{\citenamefont {Zieli\ifmmode~\acute{n}\else \'{n}\fi{}ski}\ and\
  \citenamefont {Zych}(2006)}]{Magdalena2006PRA}%
  \BibitemOpen
  \bibfield  {author} {\bibinfo {author} {\bibfnamefont {B.}~\bibnamefont
  {Zieli\ifmmode~\acute{n}\else \'{n}\fi{}ski}}\ and\ \bibinfo {author}
  {\bibfnamefont {M.}~\bibnamefont {Zych}},\ }\href
  {https://doi.org/10.1103/PhysRevA.74.034301} {\bibfield  {journal} {\bibinfo
  {journal} {Phys. Rev. A}\ }\textbf {\bibinfo {volume} {74}},\ \bibinfo
  {pages} {034301} (\bibinfo {year} {2006})}\BibitemShut {NoStop}%
\bibitem [{\citenamefont {Margolus}(2011)}]{Norman2011}%
  \BibitemOpen
  \bibfield  {author} {\bibinfo {author} {\bibfnamefont {N.}~\bibnamefont
  {Margolus}},\ }\href@noop {} {\  (\bibinfo {year} {2011})},\ \Eprint
  {https://arxiv.org/abs/arXiv:1109.4994} {arXiv:1109.4994 [quant-ph]}
  \BibitemShut {NoStop}%
\bibitem [{\citenamefont {Luo}(2004)}]{Luo2004}%
  \BibitemOpen
  \bibfield  {author} {\bibinfo {author} {\bibfnamefont {S.}~\bibnamefont
  {Luo}},\ }\href {https://doi.org/https://doi.org/10.1016/j.physd.2003.10.001}
  {\bibfield  {journal} {\bibinfo  {journal} {Physica D: Nonlinear Phenomena}\
  }\textbf {\bibinfo {volume} {189}},\ \bibinfo {pages} {1} (\bibinfo {year}
  {2004})}\BibitemShut {NoStop}%
\bibitem [{\citenamefont {Levitin}\ and\ \citenamefont
  {Toffoli}(2009)}]{Toffoli2009PRL}%
  \BibitemOpen
  \bibfield  {author} {\bibinfo {author} {\bibfnamefont {L.~B.}\ \bibnamefont
  {Levitin}}\ and\ \bibinfo {author} {\bibfnamefont {T.}~\bibnamefont
  {Toffoli}},\ }\href {https://doi.org/10.1103/PhysRevLett.103.160502}
  {\bibfield  {journal} {\bibinfo  {journal} {Phys. Rev. Lett.}\ }\textbf
  {\bibinfo {volume} {103}},\ \bibinfo {pages} {160502} (\bibinfo {year}
  {2009})}\BibitemShut {NoStop}%
\bibitem [{\citenamefont {Ness}\ \emph {et~al.}(2022)\citenamefont {Ness},
  \citenamefont {Alberti},\ and\ \citenamefont {Sagi}}]{Ness2022PRL}%
  \BibitemOpen
  \bibfield  {author} {\bibinfo {author} {\bibfnamefont {G.}~\bibnamefont
  {Ness}}, \bibinfo {author} {\bibfnamefont {A.}~\bibnamefont {Alberti}},\ and\
  \bibinfo {author} {\bibfnamefont {Y.}~\bibnamefont {Sagi}},\ }\href
  {https://doi.org/10.1103/PhysRevLett.129.140403} {\bibfield  {journal}
  {\bibinfo  {journal} {Phys. Rev. Lett.}\ }\textbf {\bibinfo {volume} {129}},\
  \bibinfo {pages} {140403} (\bibinfo {year} {2022})}\BibitemShut {NoStop}%
\bibitem [{\citenamefont {Taddei}\ \emph {et~al.}(2013)\citenamefont {Taddei},
  \citenamefont {Escher}, \citenamefont {Davidovich},\ and\ \citenamefont
  {de~Matos~Filho}}]{Taddei2013PRL}%
  \BibitemOpen
  \bibfield  {author} {\bibinfo {author} {\bibfnamefont {M.~M.}\ \bibnamefont
  {Taddei}}, \bibinfo {author} {\bibfnamefont {B.~M.}\ \bibnamefont {Escher}},
  \bibinfo {author} {\bibfnamefont {L.}~\bibnamefont {Davidovich}},\ and\
  \bibinfo {author} {\bibfnamefont {R.~L.}\ \bibnamefont {de~Matos~Filho}},\
  }\href {https://doi.org/10.1103/PhysRevLett.110.050402} {\bibfield  {journal}
  {\bibinfo  {journal} {Phys. Rev. Lett.}\ }\textbf {\bibinfo {volume} {110}},\
  \bibinfo {pages} {050402} (\bibinfo {year} {2013})}\BibitemShut {NoStop}%
\bibitem [{\citenamefont {del Campo}\ \emph {et~al.}(2013)\citenamefont {del
  Campo}, \citenamefont {Egusquiza}, \citenamefont {Plenio},\ and\
  \citenamefont {Huelga}}]{Adolfo2013PRL}%
  \BibitemOpen
  \bibfield  {author} {\bibinfo {author} {\bibfnamefont {A.}~\bibnamefont {del
  Campo}}, \bibinfo {author} {\bibfnamefont {I.~L.}\ \bibnamefont {Egusquiza}},
  \bibinfo {author} {\bibfnamefont {M.~B.}\ \bibnamefont {Plenio}},\ and\
  \bibinfo {author} {\bibfnamefont {S.~F.}\ \bibnamefont {Huelga}},\ }\href
  {https://doi.org/10.1103/PhysRevLett.110.050403} {\bibfield  {journal}
  {\bibinfo  {journal} {Phys. Rev. Lett.}\ }\textbf {\bibinfo {volume} {110}},\
  \bibinfo {pages} {050403} (\bibinfo {year} {2013})}\BibitemShut {NoStop}%
\bibitem [{\citenamefont {Deffner}\ and\ \citenamefont
  {Lutz}(2013)}]{Deffner2013PRL}%
  \BibitemOpen
  \bibfield  {author} {\bibinfo {author} {\bibfnamefont {S.}~\bibnamefont
  {Deffner}}\ and\ \bibinfo {author} {\bibfnamefont {E.}~\bibnamefont {Lutz}},\
  }\href {https://doi.org/10.1103/PhysRevLett.111.010402} {\bibfield  {journal}
  {\bibinfo  {journal} {Phys. Rev. Lett.}\ }\textbf {\bibinfo {volume} {111}},\
  \bibinfo {pages} {010402} (\bibinfo {year} {2013})}\BibitemShut {NoStop}%
\bibitem [{\citenamefont {Zhang}\ \emph {et~al.}(2014)\citenamefont {Zhang},
  \citenamefont {Han}, \citenamefont {Xia}, \citenamefont {Cao},\ and\
  \citenamefont {Fan}}]{Zhang2014}%
  \BibitemOpen
  \bibfield  {author} {\bibinfo {author} {\bibfnamefont {Y.-J.}\ \bibnamefont
  {Zhang}}, \bibinfo {author} {\bibfnamefont {W.}~\bibnamefont {Han}}, \bibinfo
  {author} {\bibfnamefont {Y.-J.}\ \bibnamefont {Xia}}, \bibinfo {author}
  {\bibfnamefont {J.-P.}\ \bibnamefont {Cao}},\ and\ \bibinfo {author}
  {\bibfnamefont {H.}~\bibnamefont {Fan}},\ }\href
  {https://doi.org/10.1038/srep04890} {\bibfield  {journal} {\bibinfo
  {journal} {Scientific Reports}\ }\textbf {\bibinfo {volume} {4}},\ \bibinfo
  {pages} {4890} (\bibinfo {year} {2014})}\BibitemShut {NoStop}%
\bibitem [{\citenamefont {Marvian}\ and\ \citenamefont
  {Lidar}(2015)}]{Lidar2015PRL}%
  \BibitemOpen
  \bibfield  {author} {\bibinfo {author} {\bibfnamefont {I.}~\bibnamefont
  {Marvian}}\ and\ \bibinfo {author} {\bibfnamefont {D.~A.}\ \bibnamefont
  {Lidar}},\ }\href {https://doi.org/10.1103/PhysRevLett.115.210402} {\bibfield
   {journal} {\bibinfo  {journal} {Phys. Rev. Lett.}\ }\textbf {\bibinfo
  {volume} {115}},\ \bibinfo {pages} {210402} (\bibinfo {year}
  {2015})}\BibitemShut {NoStop}%
\bibitem [{\citenamefont {Pires}\ \emph {et~al.}(2016)\citenamefont {Pires},
  \citenamefont {Cianciaruso}, \citenamefont {C\'eleri}, \citenamefont
  {Adesso},\ and\ \citenamefont {Soares-Pinto}}]{Pires2016PRX}%
  \BibitemOpen
  \bibfield  {author} {\bibinfo {author} {\bibfnamefont {D.~P.}\ \bibnamefont
  {Pires}}, \bibinfo {author} {\bibfnamefont {M.}~\bibnamefont {Cianciaruso}},
  \bibinfo {author} {\bibfnamefont {L.~C.}\ \bibnamefont {C\'eleri}}, \bibinfo
  {author} {\bibfnamefont {G.}~\bibnamefont {Adesso}},\ and\ \bibinfo {author}
  {\bibfnamefont {D.~O.}\ \bibnamefont {Soares-Pinto}},\ }\href
  {https://doi.org/10.1103/PhysRevX.6.021031} {\bibfield  {journal} {\bibinfo
  {journal} {Phys. Rev. X}\ }\textbf {\bibinfo {volume} {6}},\ \bibinfo {pages}
  {021031} (\bibinfo {year} {2016})}\BibitemShut {NoStop}%
\bibitem [{\citenamefont {Xu}(2016)}]{Xu_2016}%
  \BibitemOpen
  \bibfield  {author} {\bibinfo {author} {\bibfnamefont {Z.}~\bibnamefont
  {Xu}},\ }\href {https://doi.org/10.1088/1367-2630/18/7/073005} {\bibfield
  {journal} {\bibinfo  {journal} {New Journal of Physics}\ }\textbf {\bibinfo
  {volume} {18}},\ \bibinfo {pages} {073005} (\bibinfo {year}
  {2016})}\BibitemShut {NoStop}%
\bibitem [{\citenamefont {Fr\"owis}(2012)}]{Frowis2012}%
  \BibitemOpen
  \bibfield  {author} {\bibinfo {author} {\bibfnamefont {F.}~\bibnamefont
  {Fr\"owis}},\ }\href {https://doi.org/10.1103/PhysRevA.85.052127} {\bibfield
  {journal} {\bibinfo  {journal} {Phys. Rev. A}\ }\textbf {\bibinfo {volume}
  {85}},\ \bibinfo {pages} {052127} (\bibinfo {year} {2012})}\BibitemShut
  {NoStop}%
\bibitem [{\citenamefont {Xu}\ \emph {et~al.}(2014)\citenamefont {Xu},
  \citenamefont {Luo}, \citenamefont {Yang}, \citenamefont {Liu},\ and\
  \citenamefont {Zhu}}]{Zhenyu2014PRA}%
  \BibitemOpen
  \bibfield  {author} {\bibinfo {author} {\bibfnamefont {Z.}~\bibnamefont
  {Xu}}, \bibinfo {author} {\bibfnamefont {S.}~\bibnamefont {Luo}}, \bibinfo
  {author} {\bibfnamefont {W.~L.}\ \bibnamefont {Yang}}, \bibinfo {author}
  {\bibfnamefont {C.}~\bibnamefont {Liu}},\ and\ \bibinfo {author}
  {\bibfnamefont {S.}~\bibnamefont {Zhu}},\ }\href
  {https://doi.org/10.1103/PhysRevA.89.012307} {\bibfield  {journal} {\bibinfo
  {journal} {Phys. Rev. A}\ }\textbf {\bibinfo {volume} {89}},\ \bibinfo
  {pages} {012307} (\bibinfo {year} {2014})}\BibitemShut {NoStop}%
\bibitem [{\citenamefont {Zhang}\ \emph {et~al.}(2015)\citenamefont {Zhang},
  \citenamefont {Han}, \citenamefont {Xia}, \citenamefont {Cao},\ and\
  \citenamefont {Fan}}]{Zhang2015PRA}%
  \BibitemOpen
  \bibfield  {author} {\bibinfo {author} {\bibfnamefont {Y.-J.}\ \bibnamefont
  {Zhang}}, \bibinfo {author} {\bibfnamefont {W.}~\bibnamefont {Han}}, \bibinfo
  {author} {\bibfnamefont {Y.-J.}\ \bibnamefont {Xia}}, \bibinfo {author}
  {\bibfnamefont {J.-P.}\ \bibnamefont {Cao}},\ and\ \bibinfo {author}
  {\bibfnamefont {H.}~\bibnamefont {Fan}},\ }\href
  {https://doi.org/10.1103/PhysRevA.91.032112} {\bibfield  {journal} {\bibinfo
  {journal} {Phys. Rev. A}\ }\textbf {\bibinfo {volume} {91}},\ \bibinfo
  {pages} {032112} (\bibinfo {year} {2015})}\BibitemShut {NoStop}%
\bibitem [{\citenamefont {Liu}\ \emph {et~al.}(2016)\citenamefont {Liu},
  \citenamefont {Yang}, \citenamefont {An},\ and\ \citenamefont
  {Xu}}]{Liu2016PRA}%
  \BibitemOpen
  \bibfield  {author} {\bibinfo {author} {\bibfnamefont {H.-B.}\ \bibnamefont
  {Liu}}, \bibinfo {author} {\bibfnamefont {W.~L.}\ \bibnamefont {Yang}},
  \bibinfo {author} {\bibfnamefont {J.-H.}\ \bibnamefont {An}},\ and\ \bibinfo
  {author} {\bibfnamefont {Z.}~\bibnamefont {Xu}},\ }\href
  {https://doi.org/10.1103/PhysRevA.93.020105} {\bibfield  {journal} {\bibinfo
  {journal} {Phys. Rev. A}\ }\textbf {\bibinfo {volume} {93}},\ \bibinfo
  {pages} {020105(R)} (\bibinfo {year} {2016})}\BibitemShut {NoStop}%
\bibitem [{\citenamefont {Cai}\ and\ \citenamefont {Zheng}(2017)}]{Cai2017PRA}%
  \BibitemOpen
  \bibfield  {author} {\bibinfo {author} {\bibfnamefont {X.}~\bibnamefont
  {Cai}}\ and\ \bibinfo {author} {\bibfnamefont {Y.}~\bibnamefont {Zheng}},\
  }\href {https://doi.org/10.1103/PhysRevA.95.052104} {\bibfield  {journal}
  {\bibinfo  {journal} {Phys. Rev. A}\ }\textbf {\bibinfo {volume} {95}},\
  \bibinfo {pages} {052104} (\bibinfo {year} {2017})}\BibitemShut {NoStop}%
\bibitem [{\citenamefont {Wu}\ and\ \citenamefont {An}(2022)}]{An2023PRA}%
  \BibitemOpen
  \bibfield  {author} {\bibinfo {author} {\bibfnamefont {W.}~\bibnamefont
  {Wu}}\ and\ \bibinfo {author} {\bibfnamefont {J.-H.}\ \bibnamefont {An}},\
  }\href {https://doi.org/10.1103/PhysRevA.106.062438} {\bibfield  {journal}
  {\bibinfo  {journal} {Phys. Rev. A}\ }\textbf {\bibinfo {volume} {106}},\
  \bibinfo {pages} {062438} (\bibinfo {year} {2022})}\BibitemShut {NoStop}%
\bibitem [{\citenamefont {Campaioli}\ \emph {et~al.}(2018)\citenamefont
  {Campaioli}, \citenamefont {Pollock}, \citenamefont {Binder},\ and\
  \citenamefont {Modi}}]{Campaioli2018PRL}%
  \BibitemOpen
  \bibfield  {author} {\bibinfo {author} {\bibfnamefont {F.}~\bibnamefont
  {Campaioli}}, \bibinfo {author} {\bibfnamefont {F.~A.}\ \bibnamefont
  {Pollock}}, \bibinfo {author} {\bibfnamefont {F.~C.}\ \bibnamefont
  {Binder}},\ and\ \bibinfo {author} {\bibfnamefont {K.}~\bibnamefont {Modi}},\
  }\href {https://doi.org/10.1103/PhysRevLett.120.060409} {\bibfield  {journal}
  {\bibinfo  {journal} {Phys. Rev. Lett.}\ }\textbf {\bibinfo {volume} {120}},\
  \bibinfo {pages} {060409} (\bibinfo {year} {2018})}\BibitemShut {NoStop}%
\bibitem [{\citenamefont {Carlini}\ \emph {et~al.}(2006)\citenamefont
  {Carlini}, \citenamefont {Hosoya}, \citenamefont {Koike},\ and\ \citenamefont
  {Okudaira}}]{Carlini2006PRL}%
  \BibitemOpen
  \bibfield  {author} {\bibinfo {author} {\bibfnamefont {A.}~\bibnamefont
  {Carlini}}, \bibinfo {author} {\bibfnamefont {A.}~\bibnamefont {Hosoya}},
  \bibinfo {author} {\bibfnamefont {T.}~\bibnamefont {Koike}},\ and\ \bibinfo
  {author} {\bibfnamefont {Y.}~\bibnamefont {Okudaira}},\ }\href
  {https://doi.org/10.1103/PhysRevLett.96.060503} {\bibfield  {journal}
  {\bibinfo  {journal} {Phys. Rev. Lett.}\ }\textbf {\bibinfo {volume} {96}},\
  \bibinfo {pages} {060503} (\bibinfo {year} {2006})}\BibitemShut {NoStop}%
\bibitem [{\citenamefont {Caneva}\ \emph {et~al.}(2009)\citenamefont {Caneva},
  \citenamefont {Murphy}, \citenamefont {Calarco}, \citenamefont {Fazio},
  \citenamefont {Montangero}, \citenamefont {Giovannetti},\ and\ \citenamefont
  {Santoro}}]{Caneva2009PRL}%
  \BibitemOpen
  \bibfield  {author} {\bibinfo {author} {\bibfnamefont {T.}~\bibnamefont
  {Caneva}}, \bibinfo {author} {\bibfnamefont {M.}~\bibnamefont {Murphy}},
  \bibinfo {author} {\bibfnamefont {T.}~\bibnamefont {Calarco}}, \bibinfo
  {author} {\bibfnamefont {R.}~\bibnamefont {Fazio}}, \bibinfo {author}
  {\bibfnamefont {S.}~\bibnamefont {Montangero}}, \bibinfo {author}
  {\bibfnamefont {V.}~\bibnamefont {Giovannetti}},\ and\ \bibinfo {author}
  {\bibfnamefont {G.~E.}\ \bibnamefont {Santoro}},\ }\href
  {https://doi.org/10.1103/PhysRevLett.103.240501} {\bibfield  {journal}
  {\bibinfo  {journal} {Phys. Rev. Lett.}\ }\textbf {\bibinfo {volume} {103}},\
  \bibinfo {pages} {240501} (\bibinfo {year} {2009})}\BibitemShut {NoStop}%
\bibitem [{\citenamefont {Hegerfeldt}(2013)}]{Hegerfeldt2013PRL}%
  \BibitemOpen
  \bibfield  {author} {\bibinfo {author} {\bibfnamefont {G.~C.}\ \bibnamefont
  {Hegerfeldt}},\ }\href {https://doi.org/10.1103/PhysRevLett.111.260501}
  {\bibfield  {journal} {\bibinfo  {journal} {Phys. Rev. Lett.}\ }\textbf
  {\bibinfo {volume} {111}},\ \bibinfo {pages} {260501} (\bibinfo {year}
  {2013})}\BibitemShut {NoStop}%
\bibitem [{\citenamefont {Wang}\ \emph {et~al.}(2015)\citenamefont {Wang},
  \citenamefont {Allegra}, \citenamefont {Jacobs}, \citenamefont {Lloyd},
  \citenamefont {Lupo},\ and\ \citenamefont {Mohseni}}]{Wang2015PRL}%
  \BibitemOpen
  \bibfield  {author} {\bibinfo {author} {\bibfnamefont {X.}~\bibnamefont
  {Wang}}, \bibinfo {author} {\bibfnamefont {M.}~\bibnamefont {Allegra}},
  \bibinfo {author} {\bibfnamefont {K.}~\bibnamefont {Jacobs}}, \bibinfo
  {author} {\bibfnamefont {S.}~\bibnamefont {Lloyd}}, \bibinfo {author}
  {\bibfnamefont {C.}~\bibnamefont {Lupo}},\ and\ \bibinfo {author}
  {\bibfnamefont {M.}~\bibnamefont {Mohseni}},\ }\href
  {https://doi.org/10.1103/PhysRevLett.114.170501} {\bibfield  {journal}
  {\bibinfo  {journal} {Phys. Rev. Lett.}\ }\textbf {\bibinfo {volume} {114}},\
  \bibinfo {pages} {170501} (\bibinfo {year} {2015})}\BibitemShut {NoStop}%
\bibitem [{\citenamefont {Geng}\ \emph {et~al.}(2016)\citenamefont {Geng},
  \citenamefont {Wu}, \citenamefont {Wang}, \citenamefont {Xu}, \citenamefont
  {Shi}, \citenamefont {Xie}, \citenamefont {Rong},\ and\ \citenamefont
  {Du}}]{Du2016PRL}%
  \BibitemOpen
  \bibfield  {author} {\bibinfo {author} {\bibfnamefont {J.}~\bibnamefont
  {Geng}}, \bibinfo {author} {\bibfnamefont {Y.}~\bibnamefont {Wu}}, \bibinfo
  {author} {\bibfnamefont {X.}~\bibnamefont {Wang}}, \bibinfo {author}
  {\bibfnamefont {K.}~\bibnamefont {Xu}}, \bibinfo {author} {\bibfnamefont
  {F.}~\bibnamefont {Shi}}, \bibinfo {author} {\bibfnamefont {Y.}~\bibnamefont
  {Xie}}, \bibinfo {author} {\bibfnamefont {X.}~\bibnamefont {Rong}},\ and\
  \bibinfo {author} {\bibfnamefont {J.}~\bibnamefont {Du}},\ }\href
  {https://doi.org/10.1103/PhysRevLett.117.170501} {\bibfield  {journal}
  {\bibinfo  {journal} {Phys. Rev. Lett.}\ }\textbf {\bibinfo {volume} {117}},\
  \bibinfo {pages} {170501} (\bibinfo {year} {2016})}\BibitemShut {NoStop}%
\bibitem [{\citenamefont {Shiraishi}\ \emph {et~al.}(2018)\citenamefont
  {Shiraishi}, \citenamefont {Funo},\ and\ \citenamefont
  {Saito}}]{Shiraishi2018PRL}%
  \BibitemOpen
  \bibfield  {author} {\bibinfo {author} {\bibfnamefont {N.}~\bibnamefont
  {Shiraishi}}, \bibinfo {author} {\bibfnamefont {K.}~\bibnamefont {Funo}},\
  and\ \bibinfo {author} {\bibfnamefont {K.}~\bibnamefont {Saito}},\ }\href
  {https://doi.org/10.1103/PhysRevLett.121.070601} {\bibfield  {journal}
  {\bibinfo  {journal} {Phys. Rev. Lett.}\ }\textbf {\bibinfo {volume} {121}},\
  \bibinfo {pages} {070601} (\bibinfo {year} {2018})}\BibitemShut {NoStop}%
\bibitem [{\citenamefont {Garc{\'{\i}}a-Pintos}\ and\ \citenamefont {del
  Campo}(2019)}]{LuisP2019NJP}%
  \BibitemOpen
  \bibfield  {author} {\bibinfo {author} {\bibfnamefont {L.~P.}\ \bibnamefont
  {Garc{\'{\i}}a-Pintos}}\ and\ \bibinfo {author} {\bibfnamefont
  {A.}~\bibnamefont {del Campo}},\ }\href
  {https://doi.org/10.1088/1367-2630/ab099e} {\bibfield  {journal} {\bibinfo
  {journal} {New Journal of Physics}\ }\textbf {\bibinfo {volume} {21}},\
  \bibinfo {pages} {033012} (\bibinfo {year} {2019})}\BibitemShut {NoStop}%
\bibitem [{\citenamefont {Hamazaki}(2022)}]{Hamazaki2022PRXQ}%
  \BibitemOpen
  \bibfield  {author} {\bibinfo {author} {\bibfnamefont {R.}~\bibnamefont
  {Hamazaki}},\ }\href {https://doi.org/10.1103/PRXQuantum.3.020319} {\bibfield
   {journal} {\bibinfo  {journal} {PRX Quantum}\ }\textbf {\bibinfo {volume}
  {3}},\ \bibinfo {pages} {020319} (\bibinfo {year} {2022})}\BibitemShut
  {NoStop}%
\bibitem [{\citenamefont {Sun}\ and\ \citenamefont {Zheng}(2019)}]{Sun2019PRL}%
  \BibitemOpen
  \bibfield  {author} {\bibinfo {author} {\bibfnamefont {S.}~\bibnamefont
  {Sun}}\ and\ \bibinfo {author} {\bibfnamefont {Y.}~\bibnamefont {Zheng}},\
  }\href {https://doi.org/10.1103/PhysRevLett.123.180403} {\bibfield  {journal}
  {\bibinfo  {journal} {Phys. Rev. Lett.}\ }\textbf {\bibinfo {volume} {123}},\
  \bibinfo {pages} {180403} (\bibinfo {year} {2019})}\BibitemShut {NoStop}%
\bibitem [{\citenamefont {Sun}\ \emph {et~al.}(2021)\citenamefont {Sun},
  \citenamefont {Peng}, \citenamefont {Hu},\ and\ \citenamefont
  {Zheng}}]{Zheng2021PRL}%
  \BibitemOpen
  \bibfield  {author} {\bibinfo {author} {\bibfnamefont {S.}~\bibnamefont
  {Sun}}, \bibinfo {author} {\bibfnamefont {Y.}~\bibnamefont {Peng}}, \bibinfo
  {author} {\bibfnamefont {X.}~\bibnamefont {Hu}},\ and\ \bibinfo {author}
  {\bibfnamefont {Y.}~\bibnamefont {Zheng}},\ }\href
  {https://doi.org/10.1103/PhysRevLett.127.100404} {\bibfield  {journal}
  {\bibinfo  {journal} {Phys. Rev. Lett.}\ }\textbf {\bibinfo {volume} {127}},\
  \bibinfo {pages} {100404} (\bibinfo {year} {2021})}\BibitemShut {NoStop}%
\bibitem [{\citenamefont {Impens}\ \emph {et~al.}(2021)\citenamefont {Impens},
  \citenamefont {D'Angelis}, \citenamefont {Pinheiro},\ and\ \citenamefont
  {Gu\'ery-Odelin}}]{Impens2021PRA}%
  \BibitemOpen
  \bibfield  {author} {\bibinfo {author} {\bibfnamefont {F.}~\bibnamefont
  {Impens}}, \bibinfo {author} {\bibfnamefont {F.~M.}\ \bibnamefont
  {D'Angelis}}, \bibinfo {author} {\bibfnamefont {F.~A.}\ \bibnamefont
  {Pinheiro}},\ and\ \bibinfo {author} {\bibfnamefont {D.}~\bibnamefont
  {Gu\'ery-Odelin}},\ }\href {https://doi.org/10.1103/PhysRevA.104.052620}
  {\bibfield  {journal} {\bibinfo  {journal} {Phys. Rev. A}\ }\textbf {\bibinfo
  {volume} {104}},\ \bibinfo {pages} {052620} (\bibinfo {year}
  {2021})}\BibitemShut {NoStop}%
\bibitem [{\citenamefont {Fogarty}\ \emph {et~al.}(2020)\citenamefont
  {Fogarty}, \citenamefont {Deffner}, \citenamefont {Busch},\ and\
  \citenamefont {Campbell}}]{Campbell2020PRL}%
  \BibitemOpen
  \bibfield  {author} {\bibinfo {author} {\bibfnamefont {T.}~\bibnamefont
  {Fogarty}}, \bibinfo {author} {\bibfnamefont {S.}~\bibnamefont {Deffner}},
  \bibinfo {author} {\bibfnamefont {T.}~\bibnamefont {Busch}},\ and\ \bibinfo
  {author} {\bibfnamefont {S.}~\bibnamefont {Campbell}},\ }\href
  {https://doi.org/10.1103/PhysRevLett.124.110601} {\bibfield  {journal}
  {\bibinfo  {journal} {Phys. Rev. Lett.}\ }\textbf {\bibinfo {volume} {124}},\
  \bibinfo {pages} {110601} (\bibinfo {year} {2020})}\BibitemShut {NoStop}%
\bibitem [{\citenamefont {Suzuki}\ and\ \citenamefont
  {Takahashi}(2020)}]{Takahashi2020PRR}%
  \BibitemOpen
  \bibfield  {author} {\bibinfo {author} {\bibfnamefont {K.}~\bibnamefont
  {Suzuki}}\ and\ \bibinfo {author} {\bibfnamefont {K.}~\bibnamefont
  {Takahashi}},\ }\href {https://doi.org/10.1103/PhysRevResearch.2.032016}
  {\bibfield  {journal} {\bibinfo  {journal} {Phys. Rev. Research}\ }\textbf
  {\bibinfo {volume} {2}},\ \bibinfo {pages} {032016(R)} (\bibinfo {year}
  {2020})}\BibitemShut {NoStop}%
\bibitem [{\citenamefont {Garc\'{\i}a-Pintos}\ \emph
  {et~al.}(2022)\citenamefont {Garc\'{\i}a-Pintos}, \citenamefont {Nicholson},
  \citenamefont {Green}, \citenamefont {del Campo},\ and\ \citenamefont
  {Gorshkov}}]{LuisP2022PRX}%
  \BibitemOpen
  \bibfield  {author} {\bibinfo {author} {\bibfnamefont {L.~P.}\ \bibnamefont
  {Garc\'{\i}a-Pintos}}, \bibinfo {author} {\bibfnamefont {S.~B.}\ \bibnamefont
  {Nicholson}}, \bibinfo {author} {\bibfnamefont {J.~R.}\ \bibnamefont
  {Green}}, \bibinfo {author} {\bibfnamefont {A.}~\bibnamefont {del Campo}},\
  and\ \bibinfo {author} {\bibfnamefont {A.~V.}\ \bibnamefont {Gorshkov}},\
  }\href {https://doi.org/10.1103/PhysRevX.12.011038} {\bibfield  {journal}
  {\bibinfo  {journal} {Phys. Rev. X}\ }\textbf {\bibinfo {volume} {12}},\
  \bibinfo {pages} {011038} (\bibinfo {year} {2022})}\BibitemShut {NoStop}%
\bibitem [{\citenamefont {Mohan}\ and\ \citenamefont
  {Pati}(2022)}]{Mohan2022PRA}%
  \BibitemOpen
  \bibfield  {author} {\bibinfo {author} {\bibfnamefont {B.}~\bibnamefont
  {Mohan}}\ and\ \bibinfo {author} {\bibfnamefont {A.~K.}\ \bibnamefont
  {Pati}},\ }\href {https://doi.org/10.1103/PhysRevA.106.042436} {\bibfield
  {journal} {\bibinfo  {journal} {Phys. Rev. A}\ }\textbf {\bibinfo {volume}
  {106}},\ \bibinfo {pages} {042436} (\bibinfo {year} {2022})}\BibitemShut
  {NoStop}%
\bibitem [{\citenamefont {Carabba}\ \emph {et~al.}(2022)\citenamefont
  {Carabba}, \citenamefont {H{\"{o}}rnedal},\ and\ \citenamefont
  {Campo}}]{Carabba2022}%
  \BibitemOpen
  \bibfield  {author} {\bibinfo {author} {\bibfnamefont {N.}~\bibnamefont
  {Carabba}}, \bibinfo {author} {\bibfnamefont {N.}~\bibnamefont
  {H{\"{o}}rnedal}},\ and\ \bibinfo {author} {\bibfnamefont {A.~d.}\
  \bibnamefont {Campo}},\ }\href {https://doi.org/10.22331/q-2022-12-22-884}
  {\bibfield  {journal} {\bibinfo  {journal} {{Quantum}}\ }\textbf {\bibinfo
  {volume} {6}},\ \bibinfo {pages} {884} (\bibinfo {year} {2022})}\BibitemShut
  {NoStop}%
\bibitem [{\citenamefont {Van~Vu}\ and\ \citenamefont
  {Saito}(2023)}]{Tan2023PRL}%
  \BibitemOpen
  \bibfield  {author} {\bibinfo {author} {\bibfnamefont {T.}~\bibnamefont
  {Van~Vu}}\ and\ \bibinfo {author} {\bibfnamefont {K.}~\bibnamefont {Saito}},\
  }\href {https://doi.org/10.1103/PhysRevLett.130.010402} {\bibfield  {journal}
  {\bibinfo  {journal} {Phys. Rev. Lett.}\ }\textbf {\bibinfo {volume} {130}},\
  \bibinfo {pages} {010402} (\bibinfo {year} {2023})}\BibitemShut {NoStop}%
\bibitem [{\citenamefont {Lloyd}(2000)}]{lloyd2000}%
  \BibitemOpen
  \bibfield  {author} {\bibinfo {author} {\bibfnamefont {S.}~\bibnamefont
  {Lloyd}},\ }\href {https://doi.org/10.1038/35023282} {\bibfield  {journal}
  {\bibinfo  {journal} {Nature}\ }\textbf {\bibinfo {volume} {406}},\ \bibinfo
  {pages} {1047} (\bibinfo {year} {2000})}\BibitemShut {NoStop}%
\bibitem [{\citenamefont {Giovannetti}\ \emph {et~al.}(2011)\citenamefont
  {Giovannetti}, \citenamefont {Lloyd},\ and\ \citenamefont
  {Maccone}}]{Giovannetti2011}%
  \BibitemOpen
  \bibfield  {author} {\bibinfo {author} {\bibfnamefont {V.}~\bibnamefont
  {Giovannetti}}, \bibinfo {author} {\bibfnamefont {S.}~\bibnamefont {Lloyd}},\
  and\ \bibinfo {author} {\bibfnamefont {L.}~\bibnamefont {Maccone}},\ }\href
  {https://doi.org/10.1038/nphoton.2011.35} {\bibfield  {journal} {\bibinfo
  {journal} {Nature photonics}\ }\textbf {\bibinfo {volume} {5}},\ \bibinfo
  {pages} {222} (\bibinfo {year} {2011})}\BibitemShut {NoStop}%
\bibitem [{\citenamefont {Barbieri}(2022)}]{Barbieri2022PRXQuantum}%
  \BibitemOpen
  \bibfield  {author} {\bibinfo {author} {\bibfnamefont {M.}~\bibnamefont
  {Barbieri}},\ }\href {https://doi.org/10.1103/PRXQuantum.3.010202} {\bibfield
   {journal} {\bibinfo  {journal} {PRX Quantum}\ }\textbf {\bibinfo {volume}
  {3}},\ \bibinfo {pages} {010202} (\bibinfo {year} {2022})}\BibitemShut
  {NoStop}%
\bibitem [{\citenamefont {Campaioli}\ \emph {et~al.}(2017)\citenamefont
  {Campaioli}, \citenamefont {Pollock}, \citenamefont {Binder}, \citenamefont
  {C\'eleri}, \citenamefont {Goold}, \citenamefont {Vinjanampathy},\ and\
  \citenamefont {Modi}}]{Campaioli2017PRL}%
  \BibitemOpen
  \bibfield  {author} {\bibinfo {author} {\bibfnamefont {F.}~\bibnamefont
  {Campaioli}}, \bibinfo {author} {\bibfnamefont {F.~A.}\ \bibnamefont
  {Pollock}}, \bibinfo {author} {\bibfnamefont {F.~C.}\ \bibnamefont {Binder}},
  \bibinfo {author} {\bibfnamefont {L.}~\bibnamefont {C\'eleri}}, \bibinfo
  {author} {\bibfnamefont {J.}~\bibnamefont {Goold}}, \bibinfo {author}
  {\bibfnamefont {S.}~\bibnamefont {Vinjanampathy}},\ and\ \bibinfo {author}
  {\bibfnamefont {K.}~\bibnamefont {Modi}},\ }\href
  {https://doi.org/10.1103/PhysRevLett.118.150601} {\bibfield  {journal}
  {\bibinfo  {journal} {Phys. Rev. Lett.}\ }\textbf {\bibinfo {volume} {118}},\
  \bibinfo {pages} {150601} (\bibinfo {year} {2017})}\BibitemShut {NoStop}%
\bibitem [{\citenamefont {Campbell}\ and\ \citenamefont
  {Deffner}(2017)}]{Campbell2017PRL}%
  \BibitemOpen
  \bibfield  {author} {\bibinfo {author} {\bibfnamefont {S.}~\bibnamefont
  {Campbell}}\ and\ \bibinfo {author} {\bibfnamefont {S.}~\bibnamefont
  {Deffner}},\ }\href {https://doi.org/10.1103/PhysRevLett.118.100601}
  {\bibfield  {journal} {\bibinfo  {journal} {Phys. Rev. Lett.}\ }\textbf
  {\bibinfo {volume} {118}},\ \bibinfo {pages} {100601} (\bibinfo {year}
  {2017})}\BibitemShut {NoStop}%
\bibitem [{\citenamefont {Funo}\ \emph {et~al.}(2017)\citenamefont {Funo},
  \citenamefont {Zhang}, \citenamefont {Chatou}, \citenamefont {Kim},
  \citenamefont {Ueda},\ and\ \citenamefont {del Campo}}]{Funo2017PRL}%
  \BibitemOpen
  \bibfield  {author} {\bibinfo {author} {\bibfnamefont {K.}~\bibnamefont
  {Funo}}, \bibinfo {author} {\bibfnamefont {J.-N.}\ \bibnamefont {Zhang}},
  \bibinfo {author} {\bibfnamefont {C.}~\bibnamefont {Chatou}}, \bibinfo
  {author} {\bibfnamefont {K.}~\bibnamefont {Kim}}, \bibinfo {author}
  {\bibfnamefont {M.}~\bibnamefont {Ueda}},\ and\ \bibinfo {author}
  {\bibfnamefont {A.}~\bibnamefont {del Campo}},\ }\href
  {https://doi.org/10.1103/PhysRevLett.118.100602} {\bibfield  {journal}
  {\bibinfo  {journal} {Phys. Rev. Lett.}\ }\textbf {\bibinfo {volume} {118}},\
  \bibinfo {pages} {100602} (\bibinfo {year} {2017})}\BibitemShut {NoStop}%
\bibitem [{\citenamefont {Xu}\ \emph {et~al.}(2018)\citenamefont {Xu},
  \citenamefont {You}, \citenamefont {Dong}, \citenamefont {Zhang},\ and\
  \citenamefont {Yang}}]{Xu2018PRA}%
  \BibitemOpen
  \bibfield  {author} {\bibinfo {author} {\bibfnamefont {Z.}~\bibnamefont
  {Xu}}, \bibinfo {author} {\bibfnamefont {W.-L.}\ \bibnamefont {You}},
  \bibinfo {author} {\bibfnamefont {Y.-L.}\ \bibnamefont {Dong}}, \bibinfo
  {author} {\bibfnamefont {C.}~\bibnamefont {Zhang}},\ and\ \bibinfo {author}
  {\bibfnamefont {W.~L.}\ \bibnamefont {Yang}},\ }\href
  {https://doi.org/10.1103/PhysRevA.97.032115} {\bibfield  {journal} {\bibinfo
  {journal} {Phys. Rev. A}\ }\textbf {\bibinfo {volume} {97}},\ \bibinfo
  {pages} {032115} (\bibinfo {year} {2018})}\BibitemShut {NoStop}%
\bibitem [{\citenamefont {Nicholson}\ \emph {et~al.}(2020)\citenamefont
  {Nicholson}, \citenamefont {Garc{\'i}a-Pintos}, \citenamefont {del Campo},\
  and\ \citenamefont {Green}}]{Nicholson2020}%
  \BibitemOpen
  \bibfield  {author} {\bibinfo {author} {\bibfnamefont {S.~B.}\ \bibnamefont
  {Nicholson}}, \bibinfo {author} {\bibfnamefont {L.~P.}\ \bibnamefont
  {Garc{\'i}a-Pintos}}, \bibinfo {author} {\bibfnamefont {A.}~\bibnamefont {del
  Campo}},\ and\ \bibinfo {author} {\bibfnamefont {J.~R.}\ \bibnamefont
  {Green}},\ }\href {https://doi.org/10.1038/s41567-020-0981-y} {\bibfield
  {journal} {\bibinfo  {journal} {Nature Physics}\ }\textbf {\bibinfo {volume}
  {16}},\ \bibinfo {pages} {1211} (\bibinfo {year} {2020})}\BibitemShut
  {NoStop}%
\bibitem [{\citenamefont {Falasco}\ and\ \citenamefont
  {Esposito}(2020)}]{Falasco2020PRL}%
  \BibitemOpen
  \bibfield  {author} {\bibinfo {author} {\bibfnamefont {G.}~\bibnamefont
  {Falasco}}\ and\ \bibinfo {author} {\bibfnamefont {M.}~\bibnamefont
  {Esposito}},\ }\href {https://doi.org/10.1103/PhysRevLett.125.120604}
  {\bibfield  {journal} {\bibinfo  {journal} {Phys. Rev. Lett.}\ }\textbf
  {\bibinfo {volume} {125}},\ \bibinfo {pages} {120604} (\bibinfo {year}
  {2020})}\BibitemShut {NoStop}%
\bibitem [{\citenamefont {Garc\'{\i}a-Pintos}\ \emph
  {et~al.}(2020)\citenamefont {Garc\'{\i}a-Pintos}, \citenamefont {Hamma},\
  and\ \citenamefont {del Campo}}]{LuisP2020PRL}%
  \BibitemOpen
  \bibfield  {author} {\bibinfo {author} {\bibfnamefont {L.~P.}\ \bibnamefont
  {Garc\'{\i}a-Pintos}}, \bibinfo {author} {\bibfnamefont {A.}~\bibnamefont
  {Hamma}},\ and\ \bibinfo {author} {\bibfnamefont {A.}~\bibnamefont {del
  Campo}},\ }\href {https://doi.org/10.1103/PhysRevLett.125.040601} {\bibfield
  {journal} {\bibinfo  {journal} {Phys. Rev. Lett.}\ }\textbf {\bibinfo
  {volume} {125}},\ \bibinfo {pages} {040601} (\bibinfo {year}
  {2020})}\BibitemShut {NoStop}%
\bibitem [{\citenamefont {Van~Vu}\ and\ \citenamefont
  {Hasegawa}(2021)}]{VanVu2021PRL}%
  \BibitemOpen
  \bibfield  {author} {\bibinfo {author} {\bibfnamefont {T.}~\bibnamefont
  {Van~Vu}}\ and\ \bibinfo {author} {\bibfnamefont {Y.}~\bibnamefont
  {Hasegawa}},\ }\href {https://doi.org/10.1103/PhysRevLett.126.010601}
  {\bibfield  {journal} {\bibinfo  {journal} {Phys. Rev. Lett.}\ }\textbf
  {\bibinfo {volume} {126}},\ \bibinfo {pages} {010601} (\bibinfo {year}
  {2021})}\BibitemShut {NoStop}%
\bibitem [{\citenamefont {Zhen}\ \emph {et~al.}(2021)\citenamefont {Zhen},
  \citenamefont {Egloff}, \citenamefont {Modi},\ and\ \citenamefont
  {Dahlsten}}]{Dahlsten2021PRL}%
  \BibitemOpen
  \bibfield  {author} {\bibinfo {author} {\bibfnamefont {Y.-Z.}\ \bibnamefont
  {Zhen}}, \bibinfo {author} {\bibfnamefont {D.}~\bibnamefont {Egloff}},
  \bibinfo {author} {\bibfnamefont {K.}~\bibnamefont {Modi}},\ and\ \bibinfo
  {author} {\bibfnamefont {O.}~\bibnamefont {Dahlsten}},\ }\href
  {https://doi.org/10.1103/PhysRevLett.127.190602} {\bibfield  {journal}
  {\bibinfo  {journal} {Phys. Rev. Lett.}\ }\textbf {\bibinfo {volume} {127}},\
  \bibinfo {pages} {190602} (\bibinfo {year} {2021})}\BibitemShut {NoStop}%
\bibitem [{\citenamefont {Cimmarusti}\ \emph {et~al.}(2015)\citenamefont
  {Cimmarusti}, \citenamefont {Yan}, \citenamefont {Patterson}, \citenamefont
  {Corcos}, \citenamefont {Orozco},\ and\ \citenamefont
  {Deffner}}]{Cimmarusti2015}%
  \BibitemOpen
  \bibfield  {author} {\bibinfo {author} {\bibfnamefont {A.~D.}\ \bibnamefont
  {Cimmarusti}}, \bibinfo {author} {\bibfnamefont {Z.}~\bibnamefont {Yan}},
  \bibinfo {author} {\bibfnamefont {B.~D.}\ \bibnamefont {Patterson}}, \bibinfo
  {author} {\bibfnamefont {L.~P.}\ \bibnamefont {Corcos}}, \bibinfo {author}
  {\bibfnamefont {L.~A.}\ \bibnamefont {Orozco}},\ and\ \bibinfo {author}
  {\bibfnamefont {S.}~\bibnamefont {Deffner}},\ }\href
  {https://doi.org/10.1103/PhysRevLett.114.233602} {\bibfield  {journal}
  {\bibinfo  {journal} {Phys. Rev. Lett.}\ }\textbf {\bibinfo {volume} {114}},\
  \bibinfo {pages} {233602} (\bibinfo {year} {2015})}\BibitemShut {NoStop}%
\bibitem [{\citenamefont {Ness}\ \emph {et~al.}(2021)\citenamefont {Ness},
  \citenamefont {Lam}, \citenamefont {Alt}, \citenamefont {Meschede},
  \citenamefont {Sagi},\ and\ \citenamefont {Alberti}}]{Alberti2021SA}%
  \BibitemOpen
  \bibfield  {author} {\bibinfo {author} {\bibfnamefont {G.}~\bibnamefont
  {Ness}}, \bibinfo {author} {\bibfnamefont {M.~R.}\ \bibnamefont {Lam}},
  \bibinfo {author} {\bibfnamefont {W.}~\bibnamefont {Alt}}, \bibinfo {author}
  {\bibfnamefont {D.}~\bibnamefont {Meschede}}, \bibinfo {author}
  {\bibfnamefont {Y.}~\bibnamefont {Sagi}},\ and\ \bibinfo {author}
  {\bibfnamefont {A.}~\bibnamefont {Alberti}},\ }\href
  {https://doi.org/10.1126/sciadv.abj9119} {\bibfield  {journal} {\bibinfo
  {journal} {Science Advances}\ }\textbf {\bibinfo {volume} {7}},\ \bibinfo
  {pages} {eabj9119} (\bibinfo {year} {2021})}\BibitemShut {NoStop}%
\bibitem [{\citenamefont {del Campo}(2021)}]{Adolfo2021PRL}%
  \BibitemOpen
  \bibfield  {author} {\bibinfo {author} {\bibfnamefont {A.}~\bibnamefont {del
  Campo}},\ }\href {https://doi.org/10.1103/PhysRevLett.126.180603} {\bibfield
  {journal} {\bibinfo  {journal} {Phys. Rev. Lett.}\ }\textbf {\bibinfo
  {volume} {126}},\ \bibinfo {pages} {180603} (\bibinfo {year}
  {2021})}\BibitemShut {NoStop}%
\bibitem [{\citenamefont {Lam}\ \emph {et~al.}(2021)\citenamefont {Lam},
  \citenamefont {Peter}, \citenamefont {Groh}, \citenamefont {Alt},
  \citenamefont {Robens}, \citenamefont {Meschede}, \citenamefont {Negretti},
  \citenamefont {Montangero}, \citenamefont {Calarco},\ and\ \citenamefont
  {Alberti}}]{Alberti2021PRX}%
  \BibitemOpen
  \bibfield  {author} {\bibinfo {author} {\bibfnamefont {M.~R.}\ \bibnamefont
  {Lam}}, \bibinfo {author} {\bibfnamefont {N.}~\bibnamefont {Peter}}, \bibinfo
  {author} {\bibfnamefont {T.}~\bibnamefont {Groh}}, \bibinfo {author}
  {\bibfnamefont {W.}~\bibnamefont {Alt}}, \bibinfo {author} {\bibfnamefont
  {C.}~\bibnamefont {Robens}}, \bibinfo {author} {\bibfnamefont
  {D.}~\bibnamefont {Meschede}}, \bibinfo {author} {\bibfnamefont
  {A.}~\bibnamefont {Negretti}}, \bibinfo {author} {\bibfnamefont
  {S.}~\bibnamefont {Montangero}}, \bibinfo {author} {\bibfnamefont
  {T.}~\bibnamefont {Calarco}},\ and\ \bibinfo {author} {\bibfnamefont
  {A.}~\bibnamefont {Alberti}},\ }\href
  {https://doi.org/10.1103/PhysRevX.11.011035} {\bibfield  {journal} {\bibinfo
  {journal} {Phys. Rev. X}\ }\textbf {\bibinfo {volume} {11}},\ \bibinfo
  {pages} {011035} (\bibinfo {year} {2021})}\BibitemShut {NoStop}%
\bibitem [{\citenamefont {Yan}\ \emph {et~al.}(2022)\citenamefont {Yan},
  \citenamefont {Zhang}, \citenamefont {Yun}, \citenamefont {Li}, \citenamefont
  {Ding}, \citenamefont {Wei}, \citenamefont {Bu}, \citenamefont {Wang},
  \citenamefont {Chen}, \citenamefont {Su}, \citenamefont {Zhou}, \citenamefont
  {Jia}, \citenamefont {Liang},\ and\ \citenamefont {Feng}}]{Feng2022PRL}%
  \BibitemOpen
  \bibfield  {author} {\bibinfo {author} {\bibfnamefont {L.-L.}\ \bibnamefont
  {Yan}}, \bibinfo {author} {\bibfnamefont {J.-W.}\ \bibnamefont {Zhang}},
  \bibinfo {author} {\bibfnamefont {M.-R.}\ \bibnamefont {Yun}}, \bibinfo
  {author} {\bibfnamefont {J.-C.}\ \bibnamefont {Li}}, \bibinfo {author}
  {\bibfnamefont {G.-Y.}\ \bibnamefont {Ding}}, \bibinfo {author}
  {\bibfnamefont {J.-F.}\ \bibnamefont {Wei}}, \bibinfo {author} {\bibfnamefont
  {J.-T.}\ \bibnamefont {Bu}}, \bibinfo {author} {\bibfnamefont
  {B.}~\bibnamefont {Wang}}, \bibinfo {author} {\bibfnamefont {L.}~\bibnamefont
  {Chen}}, \bibinfo {author} {\bibfnamefont {S.-L.}\ \bibnamefont {Su}},
  \bibinfo {author} {\bibfnamefont {F.}~\bibnamefont {Zhou}}, \bibinfo {author}
  {\bibfnamefont {Y.}~\bibnamefont {Jia}}, \bibinfo {author} {\bibfnamefont
  {E.-J.}\ \bibnamefont {Liang}},\ and\ \bibinfo {author} {\bibfnamefont
  {M.}~\bibnamefont {Feng}},\ }\href
  {https://doi.org/10.1103/PhysRevLett.128.050603} {\bibfield  {journal}
  {\bibinfo  {journal} {Phys. Rev. Lett.}\ }\textbf {\bibinfo {volume} {128}},\
  \bibinfo {pages} {050603} (\bibinfo {year} {2022})}\BibitemShut {NoStop}%
\bibitem [{\citenamefont {Curtright}\ \emph {et~al.}(2014)\citenamefont
  {Curtright}, \citenamefont {Fairlie},\ and\ \citenamefont {Zachos}}]{bookPS}%
  \BibitemOpen
  \bibfield  {author} {\bibinfo {author} {\bibfnamefont {T.~L.}\ \bibnamefont
  {Curtright}}, \bibinfo {author} {\bibfnamefont {D.~B.}\ \bibnamefont
  {Fairlie}},\ and\ \bibinfo {author} {\bibfnamefont {C.~K.}\ \bibnamefont
  {Zachos}},\ }\href {https://doi.org/10.1142/8870} {\emph {\bibinfo {title} {A
  Concise Treatise on Quantum Mechanics in Phase Space}}}\ (\bibinfo
  {publisher} {World Scientific},\ \bibinfo {year} {2014})\BibitemShut
  {NoStop}%
\bibitem [{\citenamefont {Deffner}(2017)}]{Deffner2017NJP}%
  \BibitemOpen
  \bibfield  {author} {\bibinfo {author} {\bibfnamefont {S.}~\bibnamefont
  {Deffner}},\ }\href {https://doi.org/10.1088/1367-2630/aa83dc} {\bibfield
  {journal} {\bibinfo  {journal} {New Journal of Physics}\ }\textbf {\bibinfo
  {volume} {19}},\ \bibinfo {pages} {103018} (\bibinfo {year}
  {2017})}\BibitemShut {NoStop}%
\bibitem [{\citenamefont {Shanahan}\ \emph {et~al.}(2018)\citenamefont
  {Shanahan}, \citenamefont {Chenu}, \citenamefont {Margolus},\ and\
  \citenamefont {del Campo}}]{Shanahan2018}%
  \BibitemOpen
  \bibfield  {author} {\bibinfo {author} {\bibfnamefont {B.}~\bibnamefont
  {Shanahan}}, \bibinfo {author} {\bibfnamefont {A.}~\bibnamefont {Chenu}},
  \bibinfo {author} {\bibfnamefont {N.}~\bibnamefont {Margolus}},\ and\
  \bibinfo {author} {\bibfnamefont {A.}~\bibnamefont {del Campo}},\ }\href
  {https://doi.org/10.1103/PhysRevLett.120.070401} {\bibfield  {journal}
  {\bibinfo  {journal} {Phys. Rev. Lett.}\ }\textbf {\bibinfo {volume} {120}},\
  \bibinfo {pages} {070401} (\bibinfo {year} {2018})}\BibitemShut {NoStop}%
\bibitem [{\citenamefont {Okuyama}\ and\ \citenamefont
  {Ohzeki}(2018)}]{Okuyama2018}%
  \BibitemOpen
  \bibfield  {author} {\bibinfo {author} {\bibfnamefont {M.}~\bibnamefont
  {Okuyama}}\ and\ \bibinfo {author} {\bibfnamefont {M.}~\bibnamefont
  {Ohzeki}},\ }\href {https://doi.org/10.1103/PhysRevLett.120.070402}
  {\bibfield  {journal} {\bibinfo  {journal} {Phys. Rev. Lett.}\ }\textbf
  {\bibinfo {volume} {120}},\ \bibinfo {pages} {070402} (\bibinfo {year}
  {2018})}\BibitemShut {NoStop}%
\bibitem [{\citenamefont {Bolonek-Laso{\'{n}}}\ \emph
  {et~al.}(2021)\citenamefont {Bolonek-Laso{\'{n}}}, \citenamefont {Gonera},\
  and\ \citenamefont {Kosi{\'{n}}ski}}]{BolonekLason2021}%
  \BibitemOpen
  \bibfield  {author} {\bibinfo {author} {\bibfnamefont {K.}~\bibnamefont
  {Bolonek-Laso{\'{n}}}}, \bibinfo {author} {\bibfnamefont {J.}~\bibnamefont
  {Gonera}},\ and\ \bibinfo {author} {\bibfnamefont {P.}~\bibnamefont
  {Kosi{\'{n}}ski}},\ }\href {https://doi.org/10.22331/q-2021-06-24-482}
  {\bibfield  {journal} {\bibinfo  {journal} {{Quantum}}\ }\textbf {\bibinfo
  {volume} {5}},\ \bibinfo {pages} {482} (\bibinfo {year} {2021})}\BibitemShut
  {NoStop}%
\bibitem [{\citenamefont {Poggi}\ \emph {et~al.}(2021)\citenamefont {Poggi},
  \citenamefont {Campbell},\ and\ \citenamefont {Deffner}}]{Poggi2021PRXQ}%
  \BibitemOpen
  \bibfield  {author} {\bibinfo {author} {\bibfnamefont {P.~M.}\ \bibnamefont
  {Poggi}}, \bibinfo {author} {\bibfnamefont {S.}~\bibnamefont {Campbell}},\
  and\ \bibinfo {author} {\bibfnamefont {S.}~\bibnamefont {Deffner}},\ }\href
  {https://doi.org/10.1103/PRXQuantum.2.040349} {\bibfield  {journal} {\bibinfo
   {journal} {PRX Quantum}\ }\textbf {\bibinfo {volume} {2}},\ \bibinfo {pages}
  {040349} (\bibinfo {year} {2021})}\BibitemShut {NoStop}%
\bibitem [{\citenamefont {Wigner}(1932)}]{Wigner1932}%
  \BibitemOpen
  \bibfield  {author} {\bibinfo {author} {\bibfnamefont {E.}~\bibnamefont
  {Wigner}},\ }\href {https://doi.org/10.1103/PhysRev.40.749} {\bibfield
  {journal} {\bibinfo  {journal} {Phys. Rev.}\ }\textbf {\bibinfo {volume}
  {40}},\ \bibinfo {pages} {749} (\bibinfo {year} {1932})}\BibitemShut
  {NoStop}%
\bibitem [{\citenamefont {Glauber}(1963)}]{Glauber1963}%
  \BibitemOpen
  \bibfield  {author} {\bibinfo {author} {\bibfnamefont {R.~J.}\ \bibnamefont
  {Glauber}},\ }\href {https://doi.org/10.1103/PhysRev.131.2766} {\bibfield
  {journal} {\bibinfo  {journal} {Phys. Rev.}\ }\textbf {\bibinfo {volume}
  {131}},\ \bibinfo {pages} {2766} (\bibinfo {year} {1963})}\BibitemShut
  {NoStop}%
\bibitem [{\citenamefont {Sudarshan}(1963)}]{Sudarshan1963}%
  \BibitemOpen
  \bibfield  {author} {\bibinfo {author} {\bibfnamefont {E.~C.~G.}\
  \bibnamefont {Sudarshan}},\ }\href
  {https://doi.org/10.1103/PhysRevLett.10.277} {\bibfield  {journal} {\bibinfo
  {journal} {Phys. Rev. Lett.}\ }\textbf {\bibinfo {volume} {10}},\ \bibinfo
  {pages} {277} (\bibinfo {year} {1963})}\BibitemShut {NoStop}%
\bibitem [{\citenamefont {Leonhardt}(2010)}]{leonhardt_2010}%
  \BibitemOpen
  \bibfield  {author} {\bibinfo {author} {\bibfnamefont {U.}~\bibnamefont
  {Leonhardt}},\ }\bibinfo {title} {Quasiprobability distributions},\ in\ \href
  {https://doi.org/10.1017/CBO9780511806117.004} {\emph {\bibinfo {booktitle}
  {Essential Quantum Optics: From Quantum Measurements to Black Holes}}}\
  (\bibinfo  {publisher} {Cambridge University Press},\ \bibinfo {year}
  {2010})\ p.\ \bibinfo {pages} {63–91}\BibitemShut {NoStop}%
\bibitem [{\citenamefont {Husimi}(1940)}]{Husimi1940}%
  \BibitemOpen
  \bibfield  {author} {\bibinfo {author} {\bibfnamefont {K.}~\bibnamefont
  {Husimi}},\ }\href {https://doi.org/10.11429/ppmsj1919.22.4_264} {\bibfield
  {journal} {\bibinfo  {journal} {Proceedings of the Physico-Mathematical
  Society of Japan. 3rd Series}\ }\textbf {\bibinfo {volume} {22}},\ \bibinfo
  {pages} {264} (\bibinfo {year} {1940})}\BibitemShut {NoStop}%
\bibitem [{\citenamefont {Cahill}\ and\ \citenamefont
  {Glauber}(1969{\natexlab{a}})}]{Glauber1969}%
  \BibitemOpen
  \bibfield  {author} {\bibinfo {author} {\bibfnamefont {K.~E.}\ \bibnamefont
  {Cahill}}\ and\ \bibinfo {author} {\bibfnamefont {R.~J.}\ \bibnamefont
  {Glauber}},\ }\href {https://doi.org/10.1103/PhysRev.177.1857} {\bibfield
  {journal} {\bibinfo  {journal} {Phys. Rev.}\ }\textbf {\bibinfo {volume}
  {177}},\ \bibinfo {pages} {1857} (\bibinfo {year}
  {1969}{\natexlab{a}})}\BibitemShut {NoStop}%
\bibitem [{\citenamefont {Cahill}\ and\ \citenamefont
  {Glauber}(1969{\natexlab{b}})}]{Cahill1969}%
  \BibitemOpen
  \bibfield  {author} {\bibinfo {author} {\bibfnamefont {K.~E.}\ \bibnamefont
  {Cahill}}\ and\ \bibinfo {author} {\bibfnamefont {R.~J.}\ \bibnamefont
  {Glauber}},\ }\href {https://doi.org/10.1103/PhysRev.177.1882} {\bibfield
  {journal} {\bibinfo  {journal} {Phys. Rev.}\ }\textbf {\bibinfo {volume}
  {177}},\ \bibinfo {pages} {1882} (\bibinfo {year}
  {1969}{\natexlab{b}})}\BibitemShut {NoStop}%
\bibitem [{\citenamefont {Rundle}\ and\ \citenamefont
  {Everitt}(2021)}]{Rundle2021}%
  \BibitemOpen
  \bibfield  {author} {\bibinfo {author} {\bibfnamefont {R.~P.}\ \bibnamefont
  {Rundle}}\ and\ \bibinfo {author} {\bibfnamefont {M.~J.}\ \bibnamefont
  {Everitt}},\ }\href {https://doi.org/10.1002/qute.202100016} {\bibfield
  {journal} {\bibinfo  {journal} {Advanced Quantum Technologies}\ }\textbf
  {\bibinfo {volume} {4}},\ \bibinfo {pages} {2100016} (\bibinfo {year}
  {2021})}\BibitemShut {NoStop}%
\bibitem [{\citenamefont {Wootters}(1987)}]{Wootters1987}%
  \BibitemOpen
  \bibfield  {author} {\bibinfo {author} {\bibfnamefont {W.~K.}\ \bibnamefont
  {Wootters}},\ }\href
  {https://doi.org/https://doi.org/10.1016/0003-4916(87)90176-X} {\bibfield
  {journal} {\bibinfo  {journal} {Annals of Physics}\ }\textbf {\bibinfo
  {volume} {176}},\ \bibinfo {pages} {1} (\bibinfo {year} {1987})}\BibitemShut
  {NoStop}%
\bibitem [{\citenamefont {Heiss}\ and\ \citenamefont
  {Weigert}(2000)}]{Heiss2000PRA}%
  \BibitemOpen
  \bibfield  {author} {\bibinfo {author} {\bibfnamefont {S.}~\bibnamefont
  {Heiss}}\ and\ \bibinfo {author} {\bibfnamefont {S.}~\bibnamefont
  {Weigert}},\ }\href {https://doi.org/10.1103/PhysRevA.63.012105} {\bibfield
  {journal} {\bibinfo  {journal} {Phys. Rev. A}\ }\textbf {\bibinfo {volume}
  {63}},\ \bibinfo {pages} {012105} (\bibinfo {year} {2000})}\BibitemShut
  {NoStop}%
\bibitem [{\citenamefont {Zhu}(2016)}]{Zhu2016PRL}%
  \BibitemOpen
  \bibfield  {author} {\bibinfo {author} {\bibfnamefont {H.}~\bibnamefont
  {Zhu}},\ }\href {https://doi.org/10.1103/PhysRevLett.116.040501} {\bibfield
  {journal} {\bibinfo  {journal} {Phys. Rev. Lett.}\ }\textbf {\bibinfo
  {volume} {116}},\ \bibinfo {pages} {040501} (\bibinfo {year}
  {2016})}\BibitemShut {NoStop}%
\bibitem [{\citenamefont {Leonhardt}(1995)}]{Leonhardt2004PRL}%
  \BibitemOpen
  \bibfield  {author} {\bibinfo {author} {\bibfnamefont {U.}~\bibnamefont
  {Leonhardt}},\ }\href {https://doi.org/10.1103/PhysRevLett.74.4101}
  {\bibfield  {journal} {\bibinfo  {journal} {Phys. Rev. Lett.}\ }\textbf
  {\bibinfo {volume} {74}},\ \bibinfo {pages} {4101} (\bibinfo {year}
  {1995})}\BibitemShut {NoStop}%
\bibitem [{\citenamefont {Galv\~ao}(2005)}]{Ernesto}%
  \BibitemOpen
  \bibfield  {author} {\bibinfo {author} {\bibfnamefont {E.~F.}\ \bibnamefont
  {Galv\~ao}},\ }\href {https://doi.org/10.1103/PhysRevA.71.042302} {\bibfield
  {journal} {\bibinfo  {journal} {Phys. Rev. A}\ }\textbf {\bibinfo {volume}
  {71}},\ \bibinfo {pages} {042302} (\bibinfo {year} {2005})}\BibitemShut
  {NoStop}%
\bibitem [{\citenamefont {Veitch}\ \emph {et~al.}(2012)\citenamefont {Veitch},
  \citenamefont {Ferrie}, \citenamefont {Gross},\ and\ \citenamefont
  {Emerson}}]{Veitch_2012}%
  \BibitemOpen
  \bibfield  {author} {\bibinfo {author} {\bibfnamefont {V.}~\bibnamefont
  {Veitch}}, \bibinfo {author} {\bibfnamefont {C.}~\bibnamefont {Ferrie}},
  \bibinfo {author} {\bibfnamefont {D.}~\bibnamefont {Gross}},\ and\ \bibinfo
  {author} {\bibfnamefont {J.}~\bibnamefont {Emerson}},\ }\href
  {https://doi.org/10.1088/1367-2630/14/11/113011} {\bibfield  {journal}
  {\bibinfo  {journal} {New Journal of Physics}\ }\textbf {\bibinfo {volume}
  {14}},\ \bibinfo {pages} {113011} (\bibinfo {year} {2012})}\BibitemShut
  {NoStop}%
\bibitem [{\citenamefont {Howard}\ \emph {et~al.}(2014)\citenamefont {Howard},
  \citenamefont {Wallman}, \citenamefont {Veitch},\ and\ \citenamefont
  {Emerson}}]{Howard2014}%
  \BibitemOpen
  \bibfield  {author} {\bibinfo {author} {\bibfnamefont {M.}~\bibnamefont
  {Howard}}, \bibinfo {author} {\bibfnamefont {J.}~\bibnamefont {Wallman}},
  \bibinfo {author} {\bibfnamefont {V.}~\bibnamefont {Veitch}},\ and\ \bibinfo
  {author} {\bibfnamefont {J.}~\bibnamefont {Emerson}},\ }\href
  {https://doi.org/10.1038/nature13460} {\bibfield  {journal} {\bibinfo
  {journal} {Nature}\ }\textbf {\bibinfo {volume} {510}},\ \bibinfo {pages}
  {351} (\bibinfo {year} {2014})}\BibitemShut {NoStop}%
\bibitem [{\citenamefont {Schachenmayer}\ \emph {et~al.}(2015)\citenamefont
  {Schachenmayer}, \citenamefont {Pikovski},\ and\ \citenamefont
  {Rey}}]{Schachenmayer2015PRX}%
  \BibitemOpen
  \bibfield  {author} {\bibinfo {author} {\bibfnamefont {J.}~\bibnamefont
  {Schachenmayer}}, \bibinfo {author} {\bibfnamefont {A.}~\bibnamefont
  {Pikovski}},\ and\ \bibinfo {author} {\bibfnamefont {A.~M.}\ \bibnamefont
  {Rey}},\ }\href {https://doi.org/10.1103/PhysRevX.5.011022} {\bibfield
  {journal} {\bibinfo  {journal} {Phys. Rev. X}\ }\textbf {\bibinfo {volume}
  {5}},\ \bibinfo {pages} {011022} (\bibinfo {year} {2015})}\BibitemShut
  {NoStop}%
\bibitem [{\citenamefont {Mink}\ \emph {et~al.}(2022)\citenamefont {Mink},
  \citenamefont {Petrosyan},\ and\ \citenamefont {Fleischhauer}}]{Mink2022PRR}%
  \BibitemOpen
  \bibfield  {author} {\bibinfo {author} {\bibfnamefont {C.~D.}\ \bibnamefont
  {Mink}}, \bibinfo {author} {\bibfnamefont {D.}~\bibnamefont {Petrosyan}},\
  and\ \bibinfo {author} {\bibfnamefont {M.}~\bibnamefont {Fleischhauer}},\
  }\href {https://doi.org/10.1103/PhysRevResearch.4.043136} {\bibfield
  {journal} {\bibinfo  {journal} {Phys. Rev. Res.}\ }\textbf {\bibinfo {volume}
  {4}},\ \bibinfo {pages} {043136} (\bibinfo {year} {2022})}\BibitemShut
  {NoStop}%
\bibitem [{\citenamefont {Agarwal}(1981)}]{Agarwal1981PRA}%
  \BibitemOpen
  \bibfield  {author} {\bibinfo {author} {\bibfnamefont {G.~S.}\ \bibnamefont
  {Agarwal}},\ }\href {https://doi.org/10.1103/PhysRevA.24.2889} {\bibfield
  {journal} {\bibinfo  {journal} {Phys. Rev. A}\ }\textbf {\bibinfo {volume}
  {24}},\ \bibinfo {pages} {2889} (\bibinfo {year} {1981})}\BibitemShut
  {NoStop}%
\bibitem [{\citenamefont {Brif}\ and\ \citenamefont {Mann}(1998)}]{Brif_1998}%
  \BibitemOpen
  \bibfield  {author} {\bibinfo {author} {\bibfnamefont {C.}~\bibnamefont
  {Brif}}\ and\ \bibinfo {author} {\bibfnamefont {A.}~\bibnamefont {Mann}},\
  }\href {https://doi.org/10.1088/0305-4470/31/1/002} {\bibfield  {journal}
  {\bibinfo  {journal} {Journal of Physics A: Mathematical and General}\
  }\textbf {\bibinfo {volume} {31}},\ \bibinfo {pages} {L9} (\bibinfo {year}
  {1998})}\BibitemShut {NoStop}%
\bibitem [{\citenamefont {Brif}\ and\ \citenamefont {Mann}(1999)}]{Brif1999}%
  \BibitemOpen
  \bibfield  {author} {\bibinfo {author} {\bibfnamefont {C.}~\bibnamefont
  {Brif}}\ and\ \bibinfo {author} {\bibfnamefont {A.}~\bibnamefont {Mann}},\
  }\href {https://doi.org/10.1103/PhysRevA.59.971} {\bibfield  {journal}
  {\bibinfo  {journal} {Phys. Rev. A}\ }\textbf {\bibinfo {volume} {59}},\
  \bibinfo {pages} {971} (\bibinfo {year} {1999})}\BibitemShut {NoStop}%
\bibitem [{\citenamefont {Luis}(2004)}]{Luis2004PRA}%
  \BibitemOpen
  \bibfield  {author} {\bibinfo {author} {\bibfnamefont {A.}~\bibnamefont
  {Luis}},\ }\href {https://doi.org/10.1103/PhysRevA.69.052112} {\bibfield
  {journal} {\bibinfo  {journal} {Phys. Rev. A}\ }\textbf {\bibinfo {volume}
  {69}},\ \bibinfo {pages} {052112} (\bibinfo {year} {2004})}\BibitemShut
  {NoStop}%
\bibitem [{\citenamefont {Tilma}\ and\ \citenamefont
  {Nemoto}(2011)}]{Tilma_2012}%
  \BibitemOpen
  \bibfield  {author} {\bibinfo {author} {\bibfnamefont {T.}~\bibnamefont
  {Tilma}}\ and\ \bibinfo {author} {\bibfnamefont {K.}~\bibnamefont {Nemoto}},\
  }\href {https://doi.org/10.1088/1751-8113/45/1/015302} {\bibfield  {journal}
  {\bibinfo  {journal} {J. Phys. A: Math. Theor.}\ }\textbf {\bibinfo {volume}
  {45}},\ \bibinfo {pages} {015302} (\bibinfo {year} {2011})}\BibitemShut
  {NoStop}%
\bibitem [{\citenamefont {Tilma}\ \emph {et~al.}(2016)\citenamefont {Tilma},
  \citenamefont {Everitt}, \citenamefont {Samson}, \citenamefont {Munro},\ and\
  \citenamefont {Nemoto}}]{Tilma2016PRL}%
  \BibitemOpen
  \bibfield  {author} {\bibinfo {author} {\bibfnamefont {T.}~\bibnamefont
  {Tilma}}, \bibinfo {author} {\bibfnamefont {M.~J.}\ \bibnamefont {Everitt}},
  \bibinfo {author} {\bibfnamefont {J.~H.}\ \bibnamefont {Samson}}, \bibinfo
  {author} {\bibfnamefont {W.~J.}\ \bibnamefont {Munro}},\ and\ \bibinfo
  {author} {\bibfnamefont {K.}~\bibnamefont {Nemoto}},\ }\href
  {https://doi.org/10.1103/PhysRevLett.117.180401} {\bibfield  {journal}
  {\bibinfo  {journal} {Phys. Rev. Lett.}\ }\textbf {\bibinfo {volume} {117}},\
  \bibinfo {pages} {180401} (\bibinfo {year} {2016})}\BibitemShut {NoStop}%
\bibitem [{\citenamefont {Rundle}\ \emph {et~al.}(2019)\citenamefont {Rundle},
  \citenamefont {Tilma}, \citenamefont {Samson}, \citenamefont {Dwyer},
  \citenamefont {Bishop},\ and\ \citenamefont {Everitt}}]{Rundle2019PRA}%
  \BibitemOpen
  \bibfield  {author} {\bibinfo {author} {\bibfnamefont {R.~P.}\ \bibnamefont
  {Rundle}}, \bibinfo {author} {\bibfnamefont {T.}~\bibnamefont {Tilma}},
  \bibinfo {author} {\bibfnamefont {J.~H.}\ \bibnamefont {Samson}}, \bibinfo
  {author} {\bibfnamefont {V.~M.}\ \bibnamefont {Dwyer}}, \bibinfo {author}
  {\bibfnamefont {R.~F.}\ \bibnamefont {Bishop}},\ and\ \bibinfo {author}
  {\bibfnamefont {M.~J.}\ \bibnamefont {Everitt}},\ }\href
  {https://doi.org/10.1103/PhysRevA.99.012115} {\bibfield  {journal} {\bibinfo
  {journal} {Phys. Rev. A}\ }\textbf {\bibinfo {volume} {99}},\ \bibinfo
  {pages} {012115} (\bibinfo {year} {2019})}\BibitemShut {NoStop}%
\bibitem [{\citenamefont {Runeson}\ and\ \citenamefont
  {Richardson}(2021)}]{Runeson2021PRL}%
  \BibitemOpen
  \bibfield  {author} {\bibinfo {author} {\bibfnamefont {J.~E.}\ \bibnamefont
  {Runeson}}\ and\ \bibinfo {author} {\bibfnamefont {J.~O.}\ \bibnamefont
  {Richardson}},\ }\href {https://doi.org/10.1103/PhysRevLett.127.250403}
  {\bibfield  {journal} {\bibinfo  {journal} {Phys. Rev. Lett.}\ }\textbf
  {\bibinfo {volume} {127}},\ \bibinfo {pages} {250403} (\bibinfo {year}
  {2021})}\BibitemShut {NoStop}%
\bibitem [{\citenamefont {Stratonovich}(1957)}]{Stratonovich1957}%
  \BibitemOpen
  \bibfield  {author} {\bibinfo {author} {\bibfnamefont {R.~L.}\ \bibnamefont
  {Stratonovich}},\ }\href
  {http://jetp.ras.ru/cgi-bin/e/index/e/4/6/p891?a=list} {\bibfield  {journal}
  {\bibinfo  {journal} {Sov. Phys. JETP}\ }\textbf {\bibinfo {volume} {4}},\
  \bibinfo {pages} {891} (\bibinfo {year} {1957})}\BibitemShut {NoStop}%
\bibitem [{Not()}]{Note}%
  \BibitemOpen
  \href@noop {} {}\bibinfo {note} {Since the two components of the QSL bound
  are in dual phase spaces with opposite $s$, we may use $\mathfrak{s}:=(-s,s)$
  in the definition of $V^{\mathfrak{s}}_{\mathrm{QSL}}(t)$. For simplicity, we
  still use $V^s_{\mathrm{QSL}}(t)$ in the main text.}\BibitemShut {Stop}%
\bibitem [{\citenamefont {Gerry}\ and\ \citenamefont
  {Knight}(2004)}]{knight2004}%
  \BibitemOpen
  \bibfield  {author} {\bibinfo {author} {\bibfnamefont {C.}~\bibnamefont
  {Gerry}}\ and\ \bibinfo {author} {\bibfnamefont {P.}~\bibnamefont {Knight}},\
  }\href {https://doi.org/10.1017/CBO9780511791239} {\emph {\bibinfo {title}
  {Introductory Quantum Optics}}}\ (\bibinfo  {publisher} {Cambridge University
  Press},\ \bibinfo {year} {2004})\BibitemShut {NoStop}%
\bibitem [{\citenamefont {Klimov}\ and\ \citenamefont
  {Chumakov}(2009)}]{BookKlimov}%
  \BibitemOpen
  \bibfield  {author} {\bibinfo {author} {\bibfnamefont {A.~B.}\ \bibnamefont
  {Klimov}}\ and\ \bibinfo {author} {\bibfnamefont {S.~M.}\ \bibnamefont
  {Chumakov}},\ }\href@noop {} {\emph {\bibinfo {title} {A Group-Theoretical
  Approach to Quantum Optics: Models of Atom-Field Interactions}}}\ (\bibinfo
  {publisher} {Wiley‐VCH},\ \bibinfo {address} {Weinheim},\ \bibinfo {year}
  {2009})\BibitemShut {NoStop}%
\bibitem [{\citenamefont {Hioe}\ and\ \citenamefont
  {Eberly}(1981)}]{Eberly1981PRL}%
  \BibitemOpen
  \bibfield  {author} {\bibinfo {author} {\bibfnamefont {F.~T.}\ \bibnamefont
  {Hioe}}\ and\ \bibinfo {author} {\bibfnamefont {J.~H.}\ \bibnamefont
  {Eberly}},\ }\href {https://doi.org/10.1103/PhysRevLett.47.838} {\bibfield
  {journal} {\bibinfo  {journal} {Phys. Rev. Lett.}\ }\textbf {\bibinfo
  {volume} {47}},\ \bibinfo {pages} {838} (\bibinfo {year} {1981})}\BibitemShut
  {NoStop}%
\bibitem [{\citenamefont {Haber}(2021)}]{Haber2021}%
  \BibitemOpen
  \bibfield  {author} {\bibinfo {author} {\bibfnamefont {H.~E.}\ \bibnamefont
  {Haber}},\ }\href {https://doi.org/10.21468/SciPostPhysLectNotes.21}
  {\bibfield  {journal} {\bibinfo  {journal} {SciPost Phys. Lect. Notes}\ ,\
  \bibinfo {pages} {21}} (\bibinfo {year} {2021})}\BibitemShut {NoStop}%
\bibitem [{\citenamefont {Runeson}\ and\ \citenamefont
  {Richardson}(2020)}]{Runeson2020}%
  \BibitemOpen
  \bibfield  {author} {\bibinfo {author} {\bibfnamefont {J.~E.}\ \bibnamefont
  {Runeson}}\ and\ \bibinfo {author} {\bibfnamefont {J.~O.}\ \bibnamefont
  {Richardson}},\ }\href {https://doi.org/10.1063/1.5143412} {\bibfield
  {journal} {\bibinfo  {journal} {The Journal of Chemical Physics}\ }\textbf
  {\bibinfo {volume} {152}},\ \bibinfo {pages} {084110} (\bibinfo {year}
  {2020})}\BibitemShut {NoStop}%
\bibitem [{\citenamefont {Tian}\ \emph {et~al.}(2018)\citenamefont {Tian},
  \citenamefont {Wang}, \citenamefont {Zhang}, \citenamefont {Li},
  \citenamefont {Li},\ and\ \citenamefont {Zhang}}]{ZhangTiancai2018PRA}%
  \BibitemOpen
  \bibfield  {author} {\bibinfo {author} {\bibfnamefont {Y.}~\bibnamefont
  {Tian}}, \bibinfo {author} {\bibfnamefont {Z.}~\bibnamefont {Wang}}, \bibinfo
  {author} {\bibfnamefont {P.}~\bibnamefont {Zhang}}, \bibinfo {author}
  {\bibfnamefont {G.}~\bibnamefont {Li}}, \bibinfo {author} {\bibfnamefont
  {J.}~\bibnamefont {Li}},\ and\ \bibinfo {author} {\bibfnamefont
  {T.}~\bibnamefont {Zhang}},\ }\href
  {https://doi.org/10.1103/PhysRevA.97.013840} {\bibfield  {journal} {\bibinfo
  {journal} {Phys. Rev. A}\ }\textbf {\bibinfo {volume} {97}},\ \bibinfo
  {pages} {013840} (\bibinfo {year} {2018})}\BibitemShut {NoStop}%
\bibitem [{\citenamefont {Chen}\ \emph {et~al.}(2019)\citenamefont {Chen},
  \citenamefont {Geng}, \citenamefont {Zhou}, \citenamefont {Song},
  \citenamefont {Shen},\ and\ \citenamefont {Xu}}]{XuNanyang2019APL}%
  \BibitemOpen
  \bibfield  {author} {\bibinfo {author} {\bibfnamefont {B.}~\bibnamefont
  {Chen}}, \bibinfo {author} {\bibfnamefont {J.}~\bibnamefont {Geng}}, \bibinfo
  {author} {\bibfnamefont {F.}~\bibnamefont {Zhou}}, \bibinfo {author}
  {\bibfnamefont {L.}~\bibnamefont {Song}}, \bibinfo {author} {\bibfnamefont
  {H.}~\bibnamefont {Shen}},\ and\ \bibinfo {author} {\bibfnamefont
  {N.}~\bibnamefont {Xu}},\ }\href {https://doi.org/10.1063/1.5082878}
  {\bibfield  {journal} {\bibinfo  {journal} {Applied Physics Letters}\
  }\textbf {\bibinfo {volume} {114}},\ \bibinfo {pages} {041102} (\bibinfo
  {year} {2019})}\BibitemShut {NoStop}%
\bibitem [{\citenamefont {Lutterbach}\ and\ \citenamefont
  {Davidovich}(1997)}]{Davidovich1997PRL}%
  \BibitemOpen
  \bibfield  {author} {\bibinfo {author} {\bibfnamefont {L.~G.}\ \bibnamefont
  {Lutterbach}}\ and\ \bibinfo {author} {\bibfnamefont {L.}~\bibnamefont
  {Davidovich}},\ }\href {https://doi.org/10.1103/PhysRevLett.78.2547}
  {\bibfield  {journal} {\bibinfo  {journal} {Phys. Rev. Lett.}\ }\textbf
  {\bibinfo {volume} {78}},\ \bibinfo {pages} {2547} (\bibinfo {year}
  {1997})}\BibitemShut {NoStop}%
\bibitem [{\citenamefont {Tufarelli}\ \emph {et~al.}(2012)\citenamefont
  {Tufarelli}, \citenamefont {Ferraro}, \citenamefont {Kim},\ and\
  \citenamefont {Bose}}]{Tufarelli2012PRA}%
  \BibitemOpen
  \bibfield  {author} {\bibinfo {author} {\bibfnamefont {T.}~\bibnamefont
  {Tufarelli}}, \bibinfo {author} {\bibfnamefont {A.}~\bibnamefont {Ferraro}},
  \bibinfo {author} {\bibfnamefont {M.~S.}\ \bibnamefont {Kim}},\ and\ \bibinfo
  {author} {\bibfnamefont {S.}~\bibnamefont {Bose}},\ }\href
  {https://doi.org/10.1103/PhysRevA.85.032334} {\bibfield  {journal} {\bibinfo
  {journal} {Phys. Rev. A}\ }\textbf {\bibinfo {volume} {85}},\ \bibinfo
  {pages} {032334} (\bibinfo {year} {2012})}\BibitemShut {NoStop}%
\bibitem [{\citenamefont {Xu}\ and\ \citenamefont {del
  Campo}(2019)}]{Zhenyu2019PRL}%
  \BibitemOpen
  \bibfield  {author} {\bibinfo {author} {\bibfnamefont {Z.}~\bibnamefont
  {Xu}}\ and\ \bibinfo {author} {\bibfnamefont {A.}~\bibnamefont {del Campo}},\
  }\href {https://doi.org/10.1103/PhysRevLett.122.160602} {\bibfield  {journal}
  {\bibinfo  {journal} {Phys. Rev. Lett.}\ }\textbf {\bibinfo {volume} {122}},\
  \bibinfo {pages} {160602} (\bibinfo {year} {2019})}\BibitemShut {NoStop}%
\bibitem [{\citenamefont {Cao}\ \emph {et~al.}(2022)\citenamefont {Cao},
  \citenamefont {Xu},\ and\ \citenamefont {del Campo}}]{Zhenyu2022PRR}%
  \BibitemOpen
  \bibfield  {author} {\bibinfo {author} {\bibfnamefont {Z.}~\bibnamefont
  {Cao}}, \bibinfo {author} {\bibfnamefont {Z.}~\bibnamefont {Xu}},\ and\
  \bibinfo {author} {\bibfnamefont {A.}~\bibnamefont {del Campo}},\ }\href
  {https://doi.org/10.1103/PhysRevResearch.4.033093} {\bibfield  {journal}
  {\bibinfo  {journal} {Phys. Rev. Research}\ }\textbf {\bibinfo {volume}
  {4}},\ \bibinfo {pages} {033093} (\bibinfo {year} {2022})}\BibitemShut
  {NoStop}%
\bibitem [{\citenamefont {Vourdas}(2004)}]{Vourdas2004}%
  \BibitemOpen
  \bibfield  {author} {\bibinfo {author} {\bibfnamefont {A.}~\bibnamefont
  {Vourdas}},\ }\href {https://doi.org/10.1088/0034-4885/67/3/r03} {\bibfield
  {journal} {\bibinfo  {journal} {Reports on Progress in Physics}\ }\textbf
  {\bibinfo {volume} {67}},\ \bibinfo {pages} {267} (\bibinfo {year}
  {2004})}\BibitemShut {NoStop}%
\end{thebibliography}%

\end{document}